\documentclass[pdflatex,sn-mathphys-num]{sn-jnl}

\usepackage{graphicx}%
\usepackage{multirow}%
\usepackage{romannum}
\usepackage{wasysym}

\usepackage{amsmath,amssymb,amsfonts}%
\usepackage{amsthm}%
\usepackage{mathrsfs}%
\usepackage[title]{appendix}%
\usepackage{xcolor}%
\usepackage{textcomp}%
\usepackage{manyfoot}%
\usepackage{booktabs}%
\usepackage{algorithm}%
\usepackage{algorithmicx}%
\usepackage{algpseudocode}%
\usepackage{listings}%
\usepackage[utf8]{inputenc} 
\usepackage{newunicodechar} 
\usepackage[T1]{fontenc}
\usepackage{lmodern}
\usepackage{ragged2e}

\theoremstyle{thmstyleone}%
\theoremstyle{thmstyletwo}%

\theoremstyle{thmstylethree}%

\begin{document}

\title{Symmetry-Driven Spin Splitting in Altermagnets: An Angle-resolved photoemission spectroscopy Perspective}


\author[1,3]{\fnm{Jiayu} \sur{Liu}}\email{ljy@mail.sim.ac.cn}

\author[2]{\fnm{Xun} \sur{Ma}}\email{mx319@mail.ustc.edu.cn}

\author[2]{\fnm{Xinnuo} \sur{Zhang}}\email{zhangxinnuo@mail.ustc.edu.cn}

\author[1,3]{\fnm{Wenchuan} \sur{Jing}}\email{wcjing@mail.sim.ac.cn}

\author[1,4]{\fnm{Zhengtai} \sur{Liu}}\email{liuzt@sari.ac.cn}

\author*[2]{\fnm{Dawei} \sur{Shen}}\email{dwshen@ustc.edu.cn}

\affil[1]{\orgdiv{Shanghai Institute of Microsystem and Information Technology}, \orgname{Chinese Academy of Sciences}, \orgaddress{\city{Shanghai}, \postcode{200050}, \country{China}}}

\affil*[2]{\orgdiv{National Synchrotron Radiation Laboratory and School of Nuclear Science and Technology}, \orgname{University of Science and Technology of China}, \orgaddress{\city{Hefei}, \postcode{230026}, \country{China}}}

\affil[3]{\orgname{University of Chinese Academy of Sciences}, \orgaddress{\city{Beijing}, \postcode{100049}, \country{China}}}

\affil[4]{\orgdiv{Shanghai Synchrotron Radiation Facility}, \orgname{Shanghai Advanced Research Institute}, \orgaddress{\city{Shanghai}, \postcode{201210}, \country{China}}}


\abstract{

Altermagnetism arises from composite real-space and spin-space symmetries, combining zero net magnetization with pronounced momentum-dependent spin splitting. This review highlights the pivotal role of angle-resolved photoemission spectroscopy (ARPES)—along with its spin-resolved (SARPES) and circular-dichroism (CD-ARPES) variants, in directly visualizing the nonrelativistic band splitting and spin textures of altermagnets. Within the spin-group framework, we distinguish ferromagnetic, antiferromagnetic, and altermagnetic orders and elucidate the symmetry origin of spin polarization. We then systematically review representative systems: the debated $d$-wave prototype RuO$_2$, layered $d$-wave altermagnets KV$_2$Se$_2$O and Rb$_{1-\delta}$V$_2$Te$_2$O, and a series of $g$-wave compounds, including MnTe (domain-tunable) and CrSb (topological), together with the noncoplanar antiferromagnet MnTe$_2$ and other emerging and prospective candidates and platforms. Overall, ARPES has become a key microscope for resolving symmetry-driven spin splitting. Future advances in micro/nano-beam and \textit{in-situ} spectroscopies, combined with strain and domain engineering, heterostructure design, and the exploration of broader unconventional magnetic states, are expected to drive the joint evolution of altermagnetism and photoemission spectroscopy, paving the way for spintronic and correlated quantum research.

}




\maketitle

\section{Introduction}\label{sec1}

Altermagnetism has recently been proposed as a novel form of magnetic order, originating from composite symmetries that couple real space and spin space operations\cite{vsmejkal2022beyond, vsmejkal2022emerging,mazin2022altermagnetism,gao2025ai,mazin2021prediction,cheong2024altermagnetism}. Compared with conventional magnetic phases\cite{fritsche1998first,baltz2018antiferromagnetic,jungwirth2016antiferromagnetic}, altermagnets (AM) combine key features of both ferromagnets (FM) and antiferromagnets (AFM): on the one hand, their band structure exhibits a pronounced spin splitting reminiscent of FMs; on the other hand, the total magnetization strictly cancels out, resulting in zero net magnetization, as in AFMs. This unique combination eliminates stray-field perturbations while still providing highly efficient spin polarization, making AM an appealing platform for 
low-power spintronic devices. Both theoretical and experimental studies further suggest that AMs can host a variety of emergent phenomena, including nonrelativistic spin splitting\cite{vsmejkal2022beyond, vsmejkal2022emerging,mazin2022altermagnetism,yuan2020giant,smolyanyuk2024tool,cheong2024altermagnetism}, topological quantum states\cite{zhu2023topological,vsmejkal2023chiral}, spin--charge coupling, and unconventional transport properties\cite{vsmejkal2022anomalous,naka2019spin,feng2022anomalous,gonzalez2021efficient,bose2022tilted,ma2021multifunctional,shao2021spin,bai2022observation,das2023transport,liu2024giant,mazin2023induced,naka2022anomalous}, thereby carrying significant implications for both fundamental physics and applications\cite{vsmejkal2022beyond, vsmejkal2022emerging,bai2024altermagnetism,song2025altermagnets,rimmler2025non,ouassou2023dc,li2023majorana,zhang2024finite,dal2024antiferromagnetic,wei2024crystal,chen2024emerging,papaj2023andreev,yu2025field,wan2025high,guo2024quantum,tamang2025altermagnetism,banerjee2024altermagnetic,guo2025spin,naka2025altermagnetic,hayami2020bottom,he2023nonrelativistic,sheoran2024nonrelativistic,shim2025spin,sun2023andreev}.

To describe and classify altermagnetism, Šmejkal \textit{et al.} introduced spin group theory, which extends conventional space-group symmetry by explicitly incorporating spin degrees of freedom\cite{vsmejkal2022beyond, vsmejkal2022emerging,liu2022spin}. In this framework, the full symmetry of a system can be expressed as a product of a pure spin group $r_s$ and a nontrivial spin group $R_s$, namely $r_s \times R_s$\cite{litvin1974spin, litvin1977spin,brinkman1966theory} . The pure spin group $r_s$ acts only in spin space---for example, continuous spin rotations around a fixed axis ($C_\infty$) or a twofold rotation about the perpendicular direction ($\bar{C}_2$), ensuring that spin remains a good quantum number and defining the existence of distinct spin-up/spin-down channels. By contrast, the nontrivial spin group $R_s$ couples spin and real-space operations and is generally written as
$[C_i||C_j]$ where $C_i$ represents a spin-space operation and $C_j$ a real-space operation (such as rotation, inversion, or translation).

\begin{figure}[htbp]
\centering
\includegraphics[width=\textwidth]{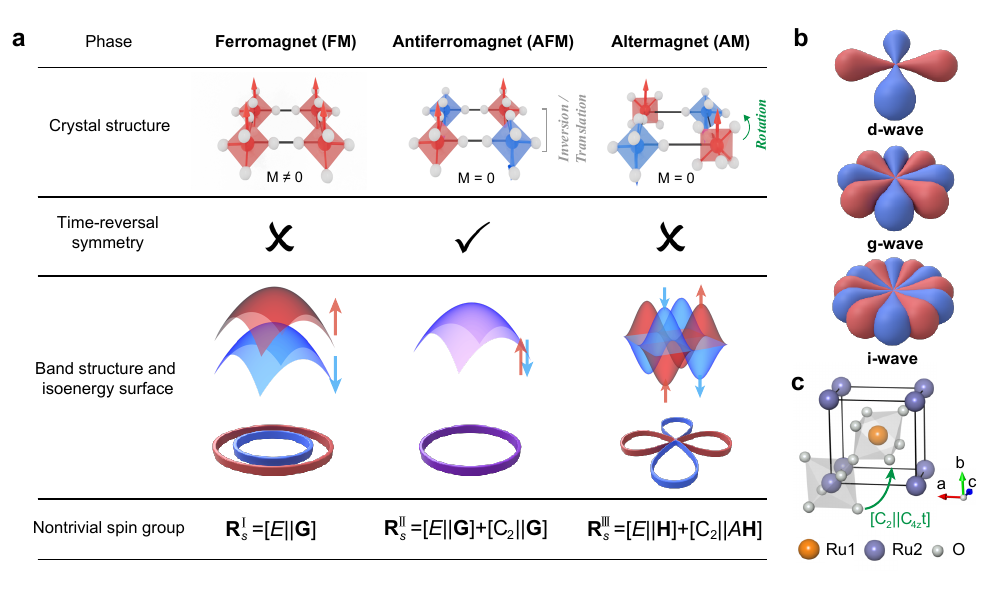} 
\caption{Fundamental magnetic phases: classification and microscopic origins. a. Comparison of the three fundamental magnetic phases—FM, AFM, and AM—in terms of their crystal structure, time-reversal symmetry, band dispersion, and nontrivial spin group. Red (blue) arrows indicate spin-up (spin-down) states\cite{vsmejkal2022beyond, vsmejkal2022emerging,mazin2022altermagnetism,cheong2024altermagnetism,gao2025ai}. b. Representative spin-polarization symmetries of $d$-, $j$-, or $i$-wave altermagnetic states\cite{vsmejkal2022beyond, vsmejkal2022emerging,tamang2025altermagnetism,bai2024altermagnetism}. c. Example of altermagnetic RuO$_2$, where the two Ru sublattices (Ru$_1$, Ru$_2$) are connected by a composite symmetry operation \([C_{2}||C_{4z}t]\).}
\end{figure}

Within this classification scheme (Fig. 1a), the three fundamental magnetic phases can be distinguished\cite{vsmejkal2022beyond, vsmejkal2022emerging,litvin1974spin, litvin1977spin}:

{
\begin{itemize}
\item \textbf{FM:} 
\(R_s^I = [E||G]\).
All spins align parallel, yielding a finite macroscopic magnetization. Here, \(E\) denotes the identity operation in spin space, corresponding to no change of spin orientation, while \(G\) denotes the crystallographic Laue group of the nonmagnetic lattice, i.e., the full set of real-space point-group symmetry operations that leave the lattice invariant. In this case, time-reversal symmetry (TRS) is fully broken\cite{slater1953ferromagnetism,eastman1978experimental,eastman1980experimental,dietl2014dilute}. 
This is because time reversal flips all spin directions, whereas the symmetry group \([E||G]\) contains no operation that can reverse spins or compensate this spin reversal through a real-space transformation. Consequently, the time-reversed state cannot be mapped back onto the original ferromagnetic configuration, and the electronic bands exhibit complete spin splitting, with spin-up and spin-down branches clearly separated (red vs.\ blue).

\item \textbf{AFM:} 
\(R_s^{II} = [E||G] + [C_2||G]\). Compared with FMs, AFMs include an additional operation that combines a twofold spin rotation (\(C_2\)) with a real-space symmetry \(G\).
The antiparallel spin alignment results in zero net magnetization. A composite antiunitary symmetry (TRS combined with translation or inversion) guarantees that spin-up and spin-down states remain degenerate in momentum space\cite{jungwirth2016antiferromagnetic,yuan2023uncovering,watson2019probing,ramazashvili2009kramers}.

\item \textbf{AM:} 
\(R_s^{III} = [E||H] + [C_2||AH]\). Here \(H\) denotes a halving subgroup of the nonmagnetic crystalline symmetry group \(G\), consisting of those real-space symmetry operations that remain valid within each magnetic sublattice once the magnetic order is taken into account, i.e., operations that do not interchange sublattices with opposite spin orientations (for example, the \(C_6\) and \(C_3\) rotations in hexagonal crystals). Accordingly, the \([E||H]\) part corresponds to spatial symmetry operations that preserve individual magnetic sublattices. These residual symmetries constrain the form of anisotropic spin-density distributions and their angular dependence, thereby permitting momentum-dependent spin polarizations such as \(d\)-, \(g\)-, or \(i\)-wave spin splittings\cite{vsmejkal2022beyond,vsmejkal2022emerging,tamang2025altermagnetism,bai2024altermagnetism} (Fig.~1b). The remaining real-space symmetry operations of the nonmagnetic group form the complementary set \(G-H \equiv AH\), which consists of operations that exchange sublattices with opposite spin orientations. The actual lifting of spin degeneracy originates from composite operations of the type \([C_2||AH]\): when such a sublattice exchange in real space is accompanied by a twofold spin rotation (\(C_2\)), time-reversal symmetry is broken at the band-structure level while zero net magnetization is preserved\cite{vsmejkal2020crystal}. This cooperative mechanism underlies the emergence of nontrivial spin splitting characteristic of altermagnets.
\end{itemize}

A prototypical example is RuO$_2$ (Fig.~1c), whose nontrivial spin group can be expressed as 
\([C_{2}||C_{4z}t]\)\cite{vsmejkal2022emerging,vsmejkal2022giant,feng2022anomalous}.
Here, \(C_{2}\) represents a \(180^{\circ}\) rotation in spin space (spin flip), 
\(C_{4z}\) a \(90^{\circ}\) rotation about the real-space \(z\)-axis, 
and \(t\) a translation mapping sublattice Ru$_1$ to Ru$_2$. 
This composite operation preserves zero net magnetization while enabling alternating-sign, 
nonrelativistic spin splitting in momentum space. 
Together with the pure spin group \(r_{s} = C_{\infty} + \bar{C}_{2}C_{\infty}\), 
this leads to the characteristic $d$-wave spin-splitting pattern in the Brillouin zone.

While the theoretical symmetry framework of altermagnetism is now established, a key experimental question remains unresolved: how can one directly verify the theoretically predicted momentum-dependent, non-relativistic spin splitting and its symmetry in real materials? Traditional magnetic characterization techniques remain essential for determining long-range magnetic order and internal field distributions. Neutron diffraction and resonant X-ray scattering accurately resolve magnetic structures and quantify small ordered moments, while muon spin relaxation/rotation ($\mu$SR) sensitively probes local magnetic fields~\cite{le2011muon,blundell2004muon}. The magneto-optical Kerr effect further enables optical probing of magnetism and magnetic domain imaging through polarization contrast. In contrast, transport measurements such as the anomalous Hall and magnetoresistance effects, though insightful, are often convoluted by contributions from Berry curvature, scattering anisotropy, and multiband transport, and thus rarely provide an unambiguous probe of altermagnetism\cite{nakatsuji2015large,qiu2000surface,fiebig2016evolution,nagaosa2010anomalous}. While these techniques remain indispensable for magnetic studies, they mainly probe real-space properties or momentum-averaged responses and cannot directly access the intrinsic momentum-space electronic structure. Consequently, momentum- and spin-resolved spectroscopies, such as ARPES and SARPES, are uniquely suited to reveal the band-level manifestations of altermagnets, including spin-split dispersions, asymmetric spin textures, and their coupling to crystal symmetry.

ARPES indeed provides precisely this capability: it directly visualizes band dispersions and Fermi surfaces, and in altermagnets, can reveal spin-split bands and the characteristic $d$-, $g$-, or $i$-wave angular patterns of spin splitting. Two advanced variants further strengthen its diagnostic power: (\romannum{1}) SARPES measures the spin polarization of individual bands, mapping momentum-space spin textures~\cite{okuda2011efficient,zhang2022angle-resolved}; (\romannum{2}) CD-ARPES exploits the chiral light–matter interaction to extract symmetry-imposed orbital and angular-momentum textures\cite{yen2024controllable,zhang2022angle-resolved}. Together, these techniques provide band-resolved, spin-resolved, and symmetry-resolved evidence that provides unambiguous signatures of altermagnetism. Recent extensions—such as bulk-sensitive soft X-ray ARPES for $k_z$ mapping and micro/nano-beam ARPES for domain-selective measurements—further mitigate artifacts arising from domain intermixing and confounding surface states~\cite{lv2019angle-resolved,okuda2011efficient,miyai2024visualization,schneider2012expanding}.

Following this rationale, the present review focuses on ARPES-based experimental evidence for altermagnetism.
We first summarize ARPES methodologies relevant to symmetry verification in altermagnetic systems (including conventional ARPES, SARPES, and CD-ARPES).
We then systematically review representative materials where band-resolved altermagnetic fingerprints have been firmly established:
the debated $d$-wave prototype RuO$_2$\cite{berlijn2017itinerant,zhu2019anomalous,feng2022anomalous,fedchenko2024observation,zhang2025probing,kessler2024absence,wenzel2025fermi,hiraishi2024nonmagnetic,plouff2025revisiting};
the layered two-dimensional $d$-wave systems KV$_2$Se$_2$O and Rb$_{1-\delta}$V$_2$Te$_2$O\cite{jiang2025metallic,zhang2025crystal};
and the $g$-wave altermagnets, including the domain-tunable hexagonal MnTe\cite{liu2024chiral,lee2024broken,krempasky2024altermagnetic,amin2024nanoscale,hajlaoui2024temperature}
and the topological compound CrSb\cite{reimers2024direct,ding2024large,biniskos2025systematic,yang2025three,li2025topological,lu2025signature}.
This discussion is followed by the noncoplanar antiferromagnet MnTe$_2$, which exhibits altermagnetic-type spin splitting\cite{zhu2024observation}.
Finally, we discuss several emerging and prospective candidate systems and symmetry-driven material platforms\cite{regmi2025altermagnetism,de2025optical,dale2024non,sakhya2025electronic,candelora2503discovery}—where experimental evidence is still developing or remains incomplete—and outline the pivotal role of ARPES in future altermagnetism research, charting a roadmap for integrating altermagnetic phenomena and materials within spintronics and correlated quantum materials.



\section{Angle-Resolved Photoemission Spectroscopy}\label{sec2}

\subsection{ARPES}

ARPES is a powerful experimental technique for probing the electronic structure of crystalline materials near the Fermi level. It played a pivotal role in condensed matter research during the late 1990s, providing the first experimental evidence for the unconventional $d$-wave symmetric superconducting gap in cuprate superconductors\cite{shen1993anomalously, zhang2022angle-resolved}.

\begin{figure}[htbp]
\centering
\includegraphics[width=\textwidth]{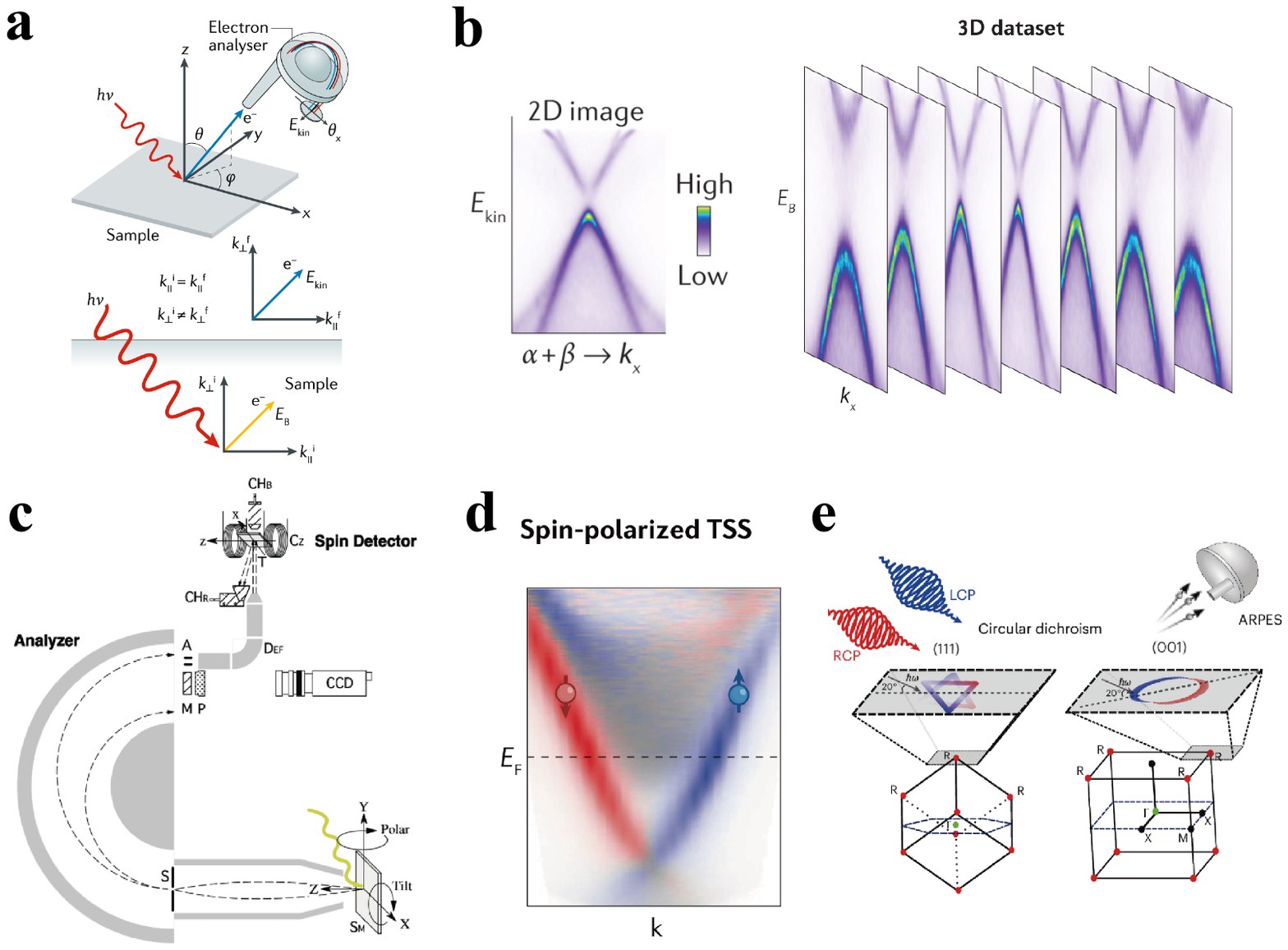} 
\caption{Experimental principles and spectral characteristics of advanced ARPES techniques. a. Schematic diagram of ARPES working principle \cite{lv2019angle-resolved}. b. Representative ARPES energy-momentum dispersion spectrum \cite{zhang2022angle-resolved}. c. Schematic illustration of SARPES apparatus with spin detection \cite{okuda2011efficient}. d. Typical spin-resolved energy distribution curves measured by SARPES \cite{zhang2022angle-resolved}. e. Operational principle of CD-ARPES using circularly polarized light \cite{yen2024controllable}.}
\end{figure}

As illustrated in Fig. 2a, ARPES measures the kinetic energy ($E_k$) and emission angle ($\theta, \phi$) of photoelectrons emitted from a sample via the photoelectric effect, thereby directly mapping the electronic energy-momentum ($E$-\textbf{k}) dispersion relations. The modern interpretation of ARPES is primarily based on the quantum mechanical one-step model. This model describes photoemission as a single, coherent quantum-mechanical transition, involving the direct optical excitation of an electron from an initial state (a Bloch wavefunction inside the solid) to a detectable free-electron final state in vacuum.

Within this framework, the photocurrent intensity \( I(\textbf{k}, E) \) measured in ARPES is given by the Fermi's Golden Rule:
\begin{equation}
I(\textbf{k}, E) \propto |M_{f,i}(\textbf{k})|^2  A(\textbf{k}, E)  f(E, T),
\end{equation}
where:
\begin{itemize}
    \item \( A(\textbf{k}, E) \) is the material's one-particle spectral function, which directly reflects the electron self-energy and its energy-momentum dispersion relation \( E(\textbf{k}) \). It is the most fundamental physical quantity probed by ARPES.
    \item \( f(E, T) \) is the Fermi-Dirac distribution function, ensuring that ARPES primarily probes the occupied states below the Fermi level \( E_F \) at a finite temperature \( T \).
    \item \( |M_{f,i}(\textbf{k})|^2 \) is the transition matrix element, governing the probability of transition from the initial state \( |i\rangle \) to the final state \( |f\rangle \).
\end{itemize}

Within the quantum mechanical one-step model framework, the photoelectron intensity \( I(\textbf{k}, E) \) measured by ARPES is directly proportional to the material's spectral function \( A(\textbf{k}, E) \). This relationship allows ARPES to directly reveal the electronic band structure \( E(\textbf{k}) \) of a solid.

The specific form of the matrix element \( M_{f,i} \) is:
\begin{equation}
M_{f,i} = \langle \psi_f(\textbf{k}) | \mathbf{A} \cdot \mathbf{p} | \psi_i(\textbf{k}) \rangle.
\end{equation}
Here, \( \mathbf{A} \) is the vector potential of the light, \( \mathbf{p} \) is the electron momentum operator, and \( \psi_i \) and \( \psi_f \) are the wavefunctions of the initial and final states, respectively. Experimental parameters, particularly photon energy and polarization, have a profound and complex influence on the measured ARPES spectra, primarily through their modulation of the transition matrix element \( |M_{f,i}(\textbf{k})|^2 \).

The variation of photon energy \( h\nu \) primarily affects the final state wavefunction \( \psi_f \), leading to two critical effects on the matrix element and the measurement: (1) Final State Effects and \( k_\perp \) Determination: The one-step model treats the final state as a Bloch wave within the crystal's periodic potential. Changing \( h\nu \) systematically alters the energy and dispersion relation of the final state. This makes it possible to probe the electron momentum perpendicular to the sample surface, \( \textbf{k}_\perp \), by scanning the photon energy. (2) Matrix Element Modulation: For certain specific \( (h\nu, \textbf{k}) \) combinations, the transition from a particular initial state \( |i\rangle \) to the final state \( |f\rangle \) may be completely suppressed (i.e., \( M_{f,i} \approx 0 \)), causing the signal from that band to vanish in the spectrum. Conversely, the same band may appear with high intensity at other photon energies. Therefore, performing photon energy-dependent measurements is a crucial practice for obtaining a complete picture of the electronic structure and avoiding the omission of key bands due to matrix element effects.

The polarization direction of the photon (the electric vector direction \( \mathbf{\hat{\epsilon}} \)) is another powerful tool for controlling the matrix element, thereby enabling the extraction of information about electron orbitals and symmetries. The transition operator \( \mathbf{A} \cdot \mathbf{p} \) implies that the photoexcitation process possesses directional selectivity. The transition probability depends on the orientation of the light's electric field vector relative to the symmetry of the initial state orbital. By carefully selecting the light's incidence geometry and polarization state ($s$-polarized, $p$-polarized, or circularly polarized), one can selectively probe electronic states with specific orbital symmetry or spatial symmetry. For instance, using linearly polarized light, it is possible to distinguish bands originating from different orbitals such as \( d_{xy} \) and \( d_{x^2-y^2} \). This is the physical basis for CD-ARPES, as illustrated in Fig. 2e, which utilizes left- and right-handed circularly polarized light to probe the chiral properties or orbital texture of electronic states.

Ultimately, ARPES transforms the spectral function \(A(\mathbf{k}, E)\)—a key physical quantity in the ``one-step model''—into a visual energy-momentum image (Fig.~2b) by measuring \(I(\mathbf{k}, E) \propto A(\mathbf{k}, E)\). The dispersion relations in the image directly map out the electronic band structure \(E(\mathbf{k})\), enabling the direct observation of electronic dispersion.

\subsection{SARPES}

SARPES represents a powerful extension of the conventional ARPES technique. While retaining the direct probing capability of the energy-momentum dispersion, it additionally resolves the spin degree of freedom of photoelectrons. This method serves as a crucial experimental tool for directly verifying and exploring the spin texture in momentum space, playing a decisive role in the study of topological quantum states and spintronics.

As illustrated in Fig.~2c, a SARPES apparatus incorporates a spin polarimeter—such as a Mott or very-low-energy electron diffraction (VLEED) detector—installed after the hemispherical energy analyzer. After analyzing the energy and emission angle (momentum) of photoelectrons, their spin polarization is measured, enabling the reconstruction of the electronic spin texture in momentum space, i.e., the momentum-dependent spin orientation.

A comprehensive spin-resolved measurement involves the following primary steps:
\begin{enumerate}
    \item \textbf{Photoexcitation and Data Acquisition}. The sample is illuminated with monochromatic light (typically linearly polarized), and the photoelectron intensity distribution at specific momentum (\textbf{k}) and binding energy ($E_B$) is collected following conventional ARPES procedures.
    
    \item \textbf{Spin Splitting and Detection}. The photoelectrons are directed into the spin polarimeter. Leveraging the spin dependence of electron-scattering processes (e.g., spin-orbit scattering), the polarimeter separates the incident electron beam into two distinct spin channels (e.g., spin-up $\uparrow$ and spin-down $\downarrow$).

    \item \textbf{Spin Polarization Calculation (Measured Asymmetry)}.
    By recording the electron counts for the two spin channels, $I_{\uparrow}$ and $I_{\downarrow}$,
    the directly measured quantity in a SARPES experiment is the asymmetry $A$ along a specific quantization axis,
    defined as
    \begin{equation}
        A = \frac{I_{\uparrow} - I_{\downarrow}}{I_{\uparrow} + I_{\downarrow}} .
        \label{eq:asymmetry}
    \end{equation}

    The asymmetry $A$ ranges from $-1$ to $+1$ and reflects the imbalance between the two spin-resolved detection channels.
    Its sign indicates the preferential spin orientation, while its magnitude is reduced relative to the true spin polarization
    due to the finite spin sensitivity of the detector.

    \item \textbf{Spin Detection and Sherman Function Correction}.
    The spin-resolved detection in SARPES relies on spin--orbit scattering processes between photoelectrons
    and atomic nuclei in the spin polarimeter.
    Because of the finite efficiency of real detectors, the measured asymmetry $A$ is smaller than the true
    spin polarization of the incident electrons.
    The physically meaningful spin polarization $P$ is therefore obtained by correcting the measured asymmetry
    using the effective Sherman function $S_{\text{eff}}$\cite{dragowski2023monte}:
    \begin{equation}
        P = \frac{A}{S_{\text{eff}}}
        = \frac{1}{S_{\text{eff}}} \cdot \frac{I_{\uparrow} - I_{\downarrow}}{I_{\uparrow} + I_{\downarrow}} .
        \label{eq:sherman_correction}
    \end{equation}

    Here, $S_{\text{eff}}$ characterizes the spin sensitivity of the detector and is typically calibrated using
    standard reference systems, such as the Rashba-split surface state of Au(111).
    Its value usually ranges between 0.1 and 0.3, depending on the detector design and calibration conditions.
    Therefore, the raw count asymmetry must be corrected by the Sherman function to obtain the true spin polarization.
    
\end{enumerate}

SARPES has not only provided direct experimental evidence for symmetry-protected helical spin textures in topological materials\cite{hsieh2009observation, souma2011direct, pan2011electronic, jozwiak2013photoelectron} (such as spin-momentum locking on topological insulator surfaces) but has also been extensively applied to quantitatively investigate spin configurations in Rashba-split systems\cite{laShell1996spin, henk2004spin--orbit, ast2007giant, bihlmayer2015focus}, magnetic materials\cite{kisker1980observation, hopster1983temperature, xu2014direct, hagiwara2016surface}, and various novel quantum materials\cite{xu2015observation, xu2016spin, gotlieb2018revealing}. It thus offers the most direct experimental window for understanding and manipulating spin-related quantum phenomena.

\subsection{CD-ARPES}

CD-ARPES combines the high-precision band-mapping capability of conventional ARPES with the sensitivity of circularly polarized light to photoemission matrix elements, which encode the symmetry properties of the initial electronic states. By comparing ARPES intensity maps acquired with left- and right-circularly polarized (LCP and RCP) photons, CD-ARPES yields a momentum- and energy-resolved dichroic response that reflects the symmetry and orbital character of electronic bands. In systems where spin–orbit coupling is relevant or time-reversal symmetry is broken, this dichroic response can further be correlated with orbital-angular-momentum textures and their associated spin structures, as schematically illustrated in Fig.~2e \cite{PhysRevB.85.195401,PhysRevLett.107.156803}.

The experimental determination of the Circular Dichroism (CD) typically involves two steps:
\begin{enumerate}
    \item Acquire two ARPES datasets using LCP and RCP light under identical experimental conditions.
    \item Evaluate the CD asymmetry, defined as:
    \begin{equation}
        \text{CD} = \frac{I_{\text{LCP}}(\textbf{k},E) - I_{\text{RCP}}(\textbf{k},E)}{I_{\text{LCP}}(\textbf{k},E) + I_{\text{RCP}}(\textbf{k},E)}.
        \label{eq:cd_asymmetry}
    \end{equation}
\end{enumerate}

The derived CD value is a normalized quantity ranging from $-1$ to $+1$ and serves as a sensitive probe of symmetry properties encoded in the photoemission matrix elements. 
Its physical interpretation can be separated into two aspects: the magnitude reflects the degree of asymmetry between left- and right-circularly polarized dipole transitions, while the sign indicates the handedness of this chiral response. 
Owing to dipole selection rules, the CD signal can, under suitable symmetry and experimental conditions, be correlated with the orbital angular momentum (OAM) character of the initial states. 
In systems where spin--orbit coupling is significant, such OAM textures may further be linked to the associated spin structures. 
Consequently, CD-ARPES provides a powerful and symmetry-sensitive tool for resolving quantum electronic states.

\begin{itemize}
    \item \textbf{Visualization of Chiral Spin Textures:} In topological materials, the strong spin-orbit coupling locks the electron's spin to its momentum and orbital character. CD-ARPES exploits this locking to indirectly visualize symmetry-protected chiral spin textures, effectively mapping the helical Dirac surface states without requiring complex Mott detectors\cite{wang2011observation, park2012chiral, kim2012spin}.
    
    \item \textbf{Identification of Orbital Selectivity:} Since different atomic orbitals (e.g., $d_{xy}$ vs.\ $d_{x^2-y^2}$) exhibit distinct symmetries, they respond differently to circularly polarized light. CD-ARPES allows for the deconvolution of these orbitally selective states, providing insight into multi-orbital physics\cite{puschnig2009reconstruction}.
    
    \item \textbf{Characterization of Inversion Symmetry Breaking:} The technique is highly sensitive to the breaking of crystal inversion symmetry. A non-zero CD signal in specific regions of the Brillouin zone can serve as an order parameter for symmetry-broken phases, helping to trace phase transitions and Berry curvature effects\cite{wallauer2021tracing}.
\end{itemize}

In the specific context of altermagnetism, CD-ARPES provides crucial symmetry-resolved information that complements spin-resolved measurements.
Rather than directly probing the nonrelativistic exchange-driven spin splitting itself, CD-ARPES is sensitive to how the photoemission matrix elements and orbital-related responses transform under the crystal point-group symmetry.
As a result, the momentum-dependent CD asymmetry offers an indirect but symmetry-selective microscopic signature of time-reversal symmetry breaking at the band-structure level, which is characteristic of altermagnetic order.
For instance, Fedchenko \emph{et al.} demonstrated that the circular dichroism observed in RuO$_2$ originates from time-reversal symmetry breaking in the bulk bands, after carefully separating geometric (CDAD) and magnetic (MCD) contributions\cite{fedchenko2024observation}, providing key spectroscopic evidence consistent with the altermagnetic scenario.

This method is particularly valuable for two reasons: First, it can be extended to other candidate materials to effectively distinguish spin splitting induced by Rashba SOC from that induced by altermagnetism. Second, for systems with small predicted splitting, CD-ARPES can identify altermagnetism by monitoring the appearance and disappearance of the CD signal across the critical temperature, even when changes in the band structure are minimal. This significantly broadens the capability of ARPES to resolve altermagnetic phases. Furthermore, recent theoretical studies indicate that driving certain two-dimensional antiferromagnetic materials with circularly polarized light can induce a transition to an altermagnetic state, representing another potential future application for CD-ARPES\cite{nn5t-kmln}.

\section{Examples of Altermagnetic Materials}\label{sec3}

Across the altermagnetic materials explored so far, two primary spin-symmetry classes have emerged: $d$-wave and $g$-wave. We begin with the prototypical $d$-wave candidate RuO$_2$, which not only ignited the modern surge of altermagnetism research but also remains one of the most debated systems to date. Although theoretical studies predict a pronounced nonrelativistic spin splitting, multiple ARPES and SARPES investigations on both bulk crystals and epitaxial films have reported band structures largely consistent with nonmagnetic calculations, leaving its magnetic ground state under continued scrutiny. Another prominent class of $d$-wave candidates is found in layered two-dimensional (2D) systems, whose exfoliability, van der Waals (vdW) stackability, and strong tunability by electric field, strain, or twist make them particularly attractive for spintronic applications. Representative examples include KV$_2$Se$_2$O and Rb$_{1-\delta}$V$_2$Te$_2$O, in which ARPES has provided unambiguous spectroscopic evidence of the characteristic altermagnetic spin splitting, directly confirming the momentum-space fingerprints of $d$-wave symmetry. In the $g$-wave family, MnTe and CrSb serve as key exemplars. For instance, ARPES studies on MnTe have revealed not only the expected symmetry-governed spin splitting but also a crucial domain-dependent symmetry reduction. In parallel, investigations into CrSb have shown that its altermagnetic order can coexist and intertwine with topological electronic features, opening avenues for exploring novel quantum phenomena. It is also instructive to consider noncoplanar antiferromagnets such as MnTe$_2$. While not strict collinear altermagnets, they also display a comparable nonrelativistic spin splitting, thereby providing a valuable baseline for distinguishing genuinely altermagnetic effects from those arising from non-collinear spin textures. Beyond these confirmed examples, several other compounds—including layered CoNb$_4$Se$_8$—have been proposed to host altermagnetism but have so far received only limited or preliminary ARPES investigation, indicating that this field is still in an early stage of experimental exploration.

To conclude this section, Table \textit{I} consolidates the current ARPES evidence and research progress across altermagnetic materials, outlining both established and yet-to-be-verified systems.

\begin{table*}[htbp]  
\centering
\caption{\justifying Summary of reported ARPES studies on representative altermagnetic materials, including their crystallographic space groups, experimental observations of spin splitting, and corresponding references. (Mn$_5$Si$_3$ has not yet been investigated by ARPES, and only transport evidence for altermagnetic behavior has been reported.)}
\label{tab1}
\renewcommand{\arraystretch}{1.2}
\begin{tabular}{|c|c|c|c|}
\hline
Materials & Space group & Spin splitting & References \\ \hline
RuO$_2$ & $P4_2/mnm$ & $\times$ & Ref.~\cite{jovic2018dirac,jovic2021momentum,uchida2020superconductivity,liu2024absence,lin2025bulk,osumi2025spin} \\ \hline
KV$_2$Se$_2$O & $P4/mmm$ & $\checked$ & Ref.~\cite{jiang2025metallic} \\ \hline
Rb$_{1-\delta}$V$_2$Te$_2$O & $P4/mmm$ & $\checked$ & Ref.~\cite{zhang2025crystal} \\ \hline
MnTe & $P6_3/mmc$ & $\checked$ & Ref.~\cite{lee2024broken,hajlaoui2024temperature,krempasky2024altermagnetic,osumi2024Observation,mcclarty2025observing,lee2025dichotomous,martuza2025itinerant,chilcote2024stoichiometry} \\ \hline
CrSb & $P6_3/mmc$ & $\checked$ & Ref.~\cite{reimers2024direct,ding2024large,yang2025three,li2025topological,lu2025signature,zeng2024observation,https://doi.org/10.1002/adma.202508977,https://doi.org/10.1002/adma.202515712,liao2025direct} \\ \hline

MnTe$_2$ & $Pa\bar{3}$ & $\checked$ & Ref.~\cite{zhu2024observation} \\ \hline
CoNb$_4$Se$_8$ & $P6_3/mmc$ & $\checked$ & Ref.~\cite{de2025optical,dale2024non,sakhya2025electronic,candelora2503discovery} \\ \hline
GdAlSi & $I4_1/md$ & $\checked$ & Ref.~\cite{PhysRevB.110.224436} \\ \hline
Mn$_5$Si$_3$ & $P6_3/mcm$ & \textbf{--}
 & Ref.~\cite{reichlova2024observation,badura2025observation} \\ \hline
Ca$_3$Ru$_2$O$_7$ & $Bb2_1m$ & $\times$ & Ref.~\cite{horio2021electronic} \\ \hline
\end{tabular}
\end{table*}

\subsection{The debate on $d$-wave altermagnetism in RuO$_2$}

Among the proposed altermagnetic candidates, RuO$_2$ stands out as one of the most promising systems, theoretically predicted to exhibit spin splittings as large as $\sim 1.4$~eV\cite{vsmejkal2022beyond, vsmejkal2022emerging,ptok2023ruthenium,sattigeri2023altermagnetic,occhialini2021local}. At the same time, its strain-induced superconductivity offers a unique opportunity to explore the interplay between the altermagnetic phase and unconventional $d$-wave superconductivity\cite{occhialini2022strain,ruf2021strain,uchida2020superconductivity}. Furthermore, theoretical studies have suggested that RuO$_2$ can host pronounced spin Hall and giant/tunneling magnetoresistance effects in antiferromagnetic tunnel junctions\cite{vsmejkal2022giant1}, while the recently observed spin splitting torque has been interpreted as supporting evidence from transport measurements, further establishing RuO$_2$ as a highly promising spin source
for spintronic applications\cite{wang2025robust,zhou2024crystal-thermal,gonzalez2021efficient,zhang2025electrical,liu2023inverse,karube2022observation,bai2023efficient,bai2022observation,shao2021spin}.

Figure 3 summarizes the historical development of research on the magnetism of RuO$_2$.
Early bulk magnetization measurements did not reveal any obvious magnetic ordering, and RuO$_2$ was long classified as a metallic Pauli paramagnet\cite{ryden1970magnetic}. A turning point came with neutron diffraction experiments, where clear scattering intensity was detected at the structurally forbidden (100) and (300) reflections of the rutile lattice, as shown in Fig. 3a. These reflections persist up to room temperature, indicating an additional ordered contribution beyond the ideal lattice\cite{berlijn2017itinerant,ahn2019antiferromagnetism}. Polarized neutron analysis further confirmed that this signal contained a magnetic component, corresponding to an effective moment of approximately $\sim$0.05~$\mu_\mathrm{B}$.
Resonant X-ray scattering likewise observed a magnetic signal at the forbidden (100) reflection, with polarization analysis showing that the intensity appeared predominantly in the $\sigma\rightarrow\pi^{\prime}$ channel while being nearly absent in the $\sigma\rightarrow\sigma^{\prime}$ channel (Fig. 3b)\cite{zhu2019anomalous}.
Transport measurements on epitaxial thin films directly reported a Berry-phase anomalous Hall effect arising from the \(\mathcal{PT}\)-symmetry
breaking of the compensated collinear order (Fig. 3c)\cite{feng2022anomalous,wang2023emergent,tschirner2023saturation}. Furthermore, Fedchenko \emph{et al.} utilized angle-resolved photoemission spectroscopy combined with magnetic circular dichroism to, by separating the CDAD from the magnetic signal MCD (Fig. 3d), demonstrate MCD$-$dominated time$-$reversal symmetry breaking at the band structure level. It is noteworthy that they subsequently rotated the sample by $180$ degrees and repeated the experiment, and the nearly unchanged CDAD signal and the reversed MCD signal further proved that the asymmetry originates from the MCD, thereby establishing a key microscopic signature of altermagnetism\cite{fedchenko2024observation}. Additional X-ray magnetic linear dichroism (XMLD) measurements revealed pronounced, orientation-dependent signals in RuO$_2$ thin films, as shown in Fig.~3e. These dichroic features vanish above the Néel temperature, providing strong evidence for antiferromagnetic ordering\cite{zhang2025probing}.

However, as research interest has grown, an increasing number of studies have challenged the magnetic ground state of RuO$_2$\cite{lin2025bulk}. Even before the concept of altermagnetism was introduced, RuO$_2$ had already been extensively investigated for its topological and superconducting properties. Band-structure studies on both single crystals and thin films consistently showed that the measured ARPES spectra are well reproduced by nonmagnetic DFT calculations, as exemplified in Fig.~3f\cite{jovic2018dirac,jovic2021momentum,uchida2020superconductivity,ruf2021strain}. Hiraishi \textit{et al.} performed $\mu$SR measurements on single crystals and found no evidence of the expected spontaneous internal field, placing an upper bound on the Ru magnetic moment at $4.8 \times 10^{-4}$~$\mu_\mathrm{B}$---nearly two orders of magnitude smaller than the $0.05$~$\mu_\mathrm{B}$ reported earlier from neutron diffraction, as shown in the left panel of Fig.~3g\cite{hiraishi2024nonmagnetic}. 
Ke{\ss}ler \textit{et al.} extended $\mu$SR experiments to both bulk and thin-film samples, establishing upper limits of $1.14 \times 10^{-4}$~$\mu_\mathrm{B}$/Ru and $7.5 \times 10^{-4}$~$\mu_\mathrm{B}$/Ru, respectively\cite{kessler2024absence,kiefer2025crystal}. 
These values are close to the detection threshold of $\mu$SR and remain far below those required to account for altermagnetic behavior. 
Moreover, their neutron diffraction measurements on single crystals, summarized in the right panel of Fig.~3g, failed to resolve a clear magnetic peak at the (100) position, detecting only weak, background-level intensity. 
They thus argued that previously reported magnetic signals likely originated from experimental artifacts (such as multiple scattering) or from sample non-stoichiometry rather than intrinsic magnetism. In line with this view, ARPES and SARPES studies by Liu \textit{et al.} demonstrated that the band structure of both single crystals and epitaxial films closely matches nonmagnetic DFT calculations, without the momentum-dependent band splitting expected for altermagnetism, as shown in Fig.~3h\cite{liu2024absence}. 
While spin-resolved measurements did reveal significant in-plane spin polarization at low-energy bulk bands, its symmetry was inconsistent with the predicted $d$-wave altermagnetic spin texture. 
Osumi \textit{et al.} subsequently proposed that such features could instead originate from orientation-dependent topological surface or interface states\cite{osumi2025spin}. Complementary experiments---including optical conductivity, time-domain terahertz spectroscopy, quantum oscillations, and thermal spin injection---have consistently characterized RuO$_2$ as a paramagnetic metal\cite{wenzel2025fermi,wu2025fermi,plouff2025revisiting,wang2025robust}. 
As shown in Fig.~3i, the left panel presents optical conductivity measurements, while the right panel shows quantum oscillation data; in both cases, the experimentally extracted spectra and oscillation frequencies are more consistently reproduced by nonmagnetic electronic structure calculations.

Amid the ongoing experimental contradictions, several theoretical studies have provided key insights.
It has been proposed that RuO$_2$ lies intrinsically close to a Landau--Pomeranchuk instability\cite{ahn2019antiferromagnetism}, which renders its ground state extremely sensitive to control parameters such as the Hubbard U, doping, and epitaxial strain. Calculations by Qian \textit{et al.} demonstrated that moderate hole doping combined with in-plane tensile strain can substantially reduce the critical threshold for stabilizing antiferromagnetic order, thereby rationalizing part of the discrepancy between experimental and theoretical results\cite{qian2025fragile}.
Complementary DFT+U analyses by Smolyanyuk \textit{et al.} revealed that stoichiometric RuO$_2$ remains nonmagnetic for $\mathrm{U}_{\mathrm{eff}} < 1.06\,\mathrm{eV}$, but once this threshold is exceeded, the Ru moment abruptly jumps to $\sim$0.5~$\mu_\mathrm{B}$\cite{smolyanyuk2024fragility,smolyanyuk2025origin}.
This finding implies that realistic stoichiometric samples, which are generally described by smaller effective U values, are more likely to remain nonmagnetic.
Additional theoretical work further indicated that factors such as Ru vacancies or hole doping can significantly lower the energetic barrier for magnetic ordering. In thin films, meanwhile, strong interlayer relaxation has been proposed to effectively introduce Hubbard-U--like effects that facilitate the emergence of altermagnetism.
Collectively, these findings underscore that the magnetic ground state of RuO$_2$ is highly contingent upon stoichiometry, epitaxial strain, and growth conditions, offering a plausible explanation for the sharply divergent experimental observations\cite{jeong2024altermagnetic,brahimi2024confinement,qian2025fragile,smolyanyuk2024fragility,liu2024absence}.

\begin{figure}[htbp]
\centering
\includegraphics[width=\textwidth]{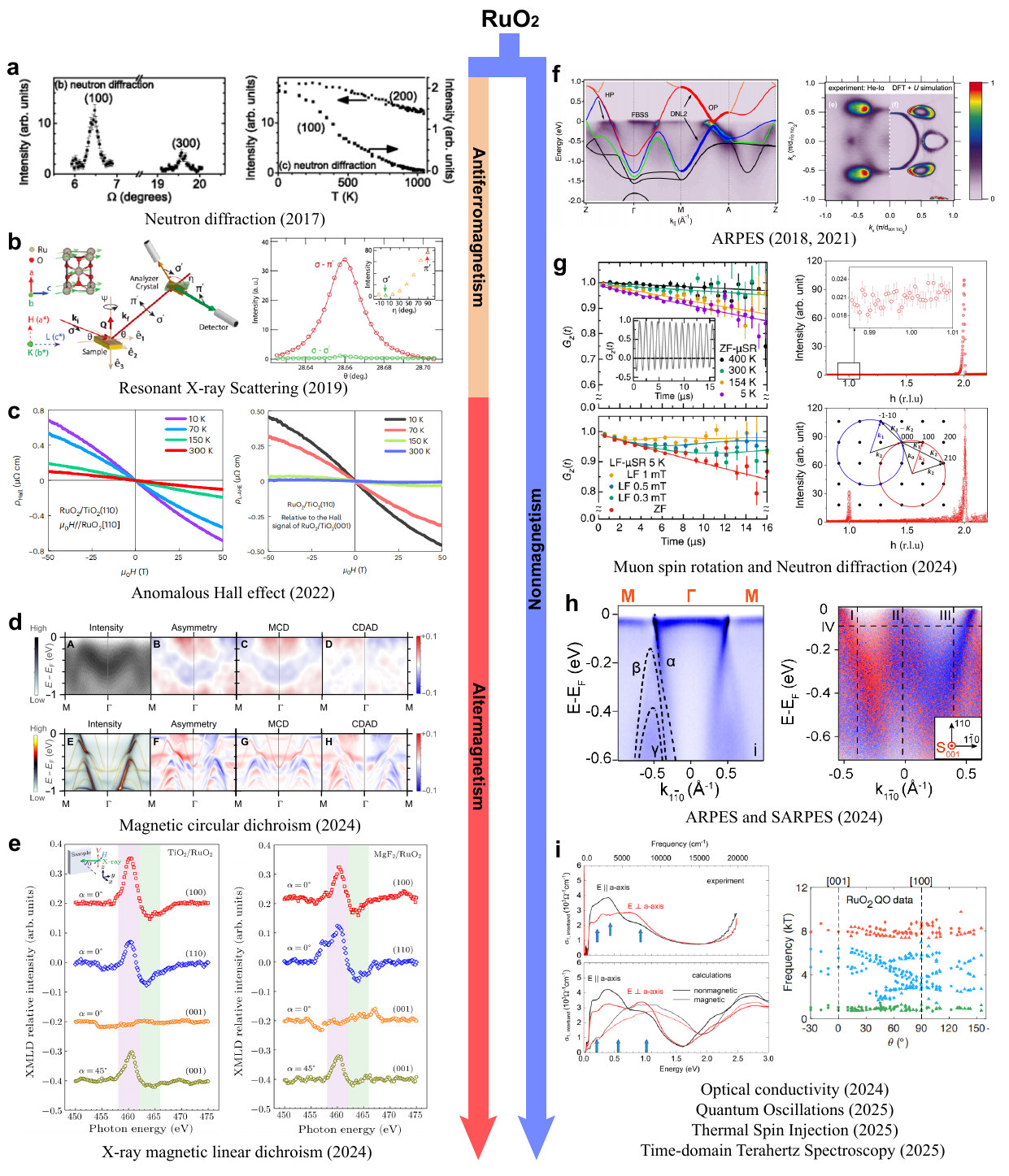}
\caption{Summary of the major advances in the magnetic studies of RuO$_2$\cite{berlijn2017itinerant,zhu2019anomalous,feng2022anomalous,fedchenko2024observation,zhang2025probing,kessler2024absence,wenzel2025fermi,hiraishi2024nonmagnetic,plouff2025revisiting,wu2025fermi,wang2025robust,uchida2020superconductivity,ruf2021strain,liu2024absence}.
a. Neutron diffraction revealing scattering intensity at structurally forbidden reflections.
b. Resonant X-ray scattering showing polarization-dependent intensity indicative of a magnetic contribution.
c. Transport measurements on epitaxial thin films revealing an anomalous Hall effect.
d. ARPES combined with magnetic circular dichroism, providing momentum-resolved evidence for time-reversal symmetry breaking at the band-structure level.
e. X-ray magnetic linear dichroism measurements showing pronounced, orientation-dependent dichroic signals.
f. ARPES results of single crystals and thin films, in good agreement with nonmagnetic DFT calculations.
g. $\mu$SR measurements (left) setting stringent upper bounds on the Ru magnetic moment, and neutron diffraction measurements (right) showing no clear magnetic peak at (100).
h. ARPES and SARPES results showing the absence of the momentum-dependent band splitting expected for altermagnetism.
i. Complementary probes of the electronic ground state, including optical conductivity (left) and quantum oscillations (right), whose data are better reproduced by nonmagnetic calculations.
}
\end{figure}

\subsection{2D altermagnets: KV$_2$Se$_2$O and Rb$_{1-\delta}$V$_2$Te$_2$O}

Exploring altermagnetism in layered two-dimensional compounds provides an ideal platform for advancing spintronics and correlated electron physics. Compared with three-dimensional altermagnets, layered systems can be exfoliated, integrated into van der Waals heterostructures, and respond more efficiently to external tuning such as electric fields, strain, or twist engineering\cite{lin2018structure,de2014ubiquitous,cao2018unconventional,flensberg2021engineered,zhao2020strain}. These attributes offer clear advantages for high-density, low-power spintronic devices\cite{jungwirth2016antiferromagnetic,baltz2018antiferromagnetic,qin2023room}.

Representative examples include KV$_2$Se$_2$O and Rb$_{1-\delta}$V$_2$Te$_2$O, both adopting a layered tetragonal ($P4/mmm$) structure\cite{jiang2025metallic,zhang2025crystal}. In KV$_2$Se$_2$O, the V$_2$O layers host an altermagnetic order connected by [$C_2||C_{4z}$] spin-group symmetry (Fig.~4a)\cite{jiang2025metallic,lin2018structure}. Nuclear magnetic resonance (NMR) spectra (Fig.~4e) showed a single quadrupole-split peak at high temperature, which split into two below 100~K, indicating the emergence of a spin-density-wave (SDW) phase with two distinct magnetic moment magnitudes\cite{jiang2025metallic,kikuchi1994antiferromagnetic,ohama1999mixed}. Additional field-induced splitting confirmed antiparallel spin alignment along $c$, consistent with a collinear compensated order.

ARPES systematically mapped the band and Fermi surface (Figs.~4b–d)\cite{jiang2025metallic}. Clear spin splitting appears along $\Gamma$–$X$ and $X$–$M$ but remains degenerate along $\Gamma$–$M$, precisely matching the $d$-wave symmetry predictions. Calculations predict a maximum splitting of $\sim$1.6~eV at the $X/Y$ points—the largest among reported altermagnets. The observed Fermi surface (Fig.~4c) displays quasi-1D $\alpha$ and $\gamma$ pockets consistent with theory, which provide the necessary Fermi surface nesting conditions for the low-temperature SDW formation. By sampling higher Brillouin-zone cuts and rotating the crystal (Fig.~4d), in-plane spin contrast was enhanced, allowing separation of spin-up and spin-down MDCs (Fig.~4f). The spin polarization reverses antisymmetrically across $\Gamma$–$M$, definitively confirming a $d$-wave spin texture in KV$_2$Se$_2$O.

Upon entering the SDW phase, ARPES spectra revealed strong band folding (Fig.~4g)\cite{jiang2025metallic}. Near $E_F$, folded $\alpha'$ and $\gamma'$ bands hybridized with the originals, opening uniform nesting gaps connected by $q=(\pi/a,\pi/a)$. Temperature-dependent data (Fig.~4h) showed the gap persisting above the SDW transition with vanishing coherence peaks—indicative of pseudogap-like behavior. Combining ARPES and NMR, the SDW magnetic structure (Fig.~4i) was deduced as two antiparallel sublattices linked by $[C_2||M_{1\bar{1}0}]$. Crucially, this SDW phase emerges in addition to, and coexists with, the underlying altermagnetic order, as the band folding builds upon the already spin-split high-temperature electronic structure. KV$_2$Se$_2$O thus emerges as a remarkable platform exhibiting coexisting $d$-wave spin splitting, FS-nesting-driven SDW, and pseudogap phenomena, revealing a unique interplay between altermagnetism and electronic correlations.

\begin{figure}[htbp]
\centering
\includegraphics[width=\textwidth]{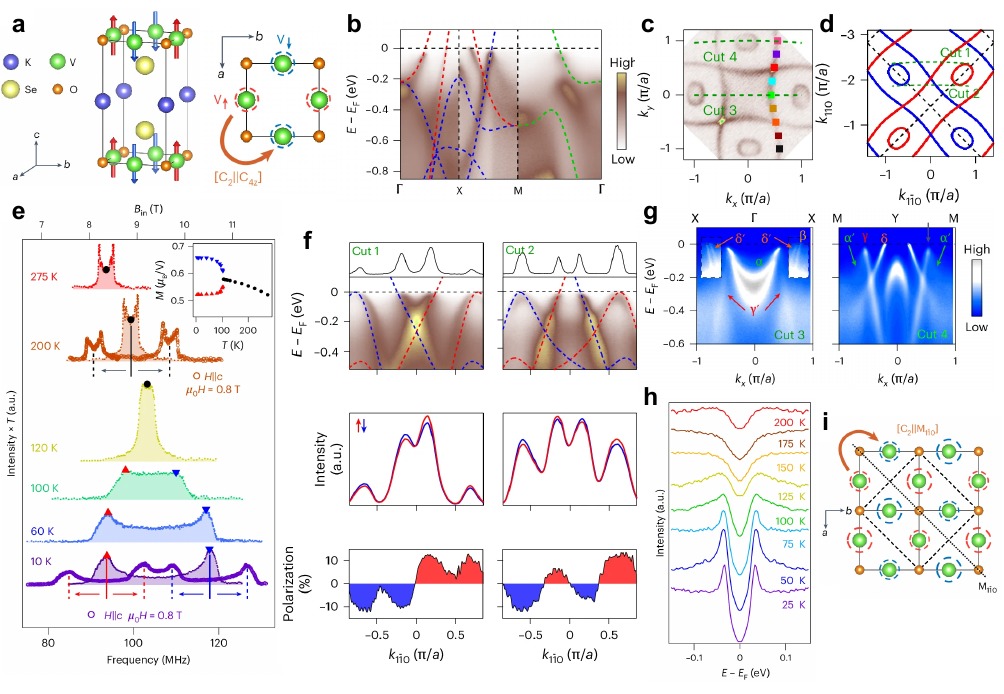} 
\caption{Electronic and magnetic properties of the layered altermagnet KV$_2$Se$_2$O\cite{jiang2025metallic}. a. Crystal structure of KV$_2$Se$_2$O showing the V$_2$O planes with alternating spin orientations connected by the spin-group symmetry [$C_2||C_{4z}$]. b. Comparison between experimental ARPES dispersions and theoretical band calculations along high-symmetry directions. c,d. Fermi surface mappings and corresponding calculations. e. Temperature-dependent $^{51}$V NMR spectra under $\mu_0H=0.8$~T. f. Band dispersions and SARPES momentum distribution curves (MDCs) and spin polarizations along cuts 1 and 2. g. Band dispersions along cuts~3 and~4, where $\alpha$, $\beta$, $\gamma$, and $\delta$ denote the original bands, while $\alpha'$, $\gamma'$, and $\delta'$ represent their folded counterparts. h. Temperature-dependent energy distribution curves (EDCs) showing that the SDW gap persists above the transition temperature. i. Schematic magnetic structure of the SDW phase. 
The antiparallel spin sublattices are linked by the [$C_2||M_{1\bar{1}0}$] symmetry, maintaining compensated magnetization while preserving the underlying $d$-wave altermagnetic order.}
\end{figure}

Distinct from KV$_2$Se$_2$O, Rb$_{1-\delta}$V$_2$Te$_2$O exhibits a crystalline symmetry that couples opposite-spin magnetic sublattices in real space and spin–momentum in reciprocal space, producing a novel C-paired spin–valley locking (SVL), as illustrated in Fig.~5a(3)\cite{zhang2025crystal}. Unlike the T-paired SVL of conventional dichalcogenides, which relies on SOC and time-reversal symmetry, Type-II C-paired SVL originates from mirror ($M_1$, $M_2$) and fourfold ($C_{4z}$) symmetries of magnetic sublattices without requiring SOC. Structurally, Rb$_{1-\delta}$V$_2$Te$_2$O consists of stacked $\mathrm{Rb}$–$\mathrm{Te}$–$\mathrm{V_2O}$–$\mathrm{Te}$ layers, forming a van der Waals-type architecture\cite{ablimit2018weak,ablimit2018v2te2o}. Magnetically, it hosts an antiferromagnetic configuration in-plane and ferromagnetic coupling along $c$. This symmetry lifts spin degeneracy and induces alternating valley spin polarization even in the absence of SOC.

In terms of spin transport, C-paired SVL can support a variety of physical responses\cite{ma2021multifunctional,wu2007fermi,naka2019spin,gonzalez2021efficient}. If each magnetic sublattice exhibited $C_n$ ($n\geq3$) symmetry, conductivity would become isotropic and the spin current non-polarized (type I). By contrast, type II systems maintain anisotropic conductivity within each
spin channel, thereby allowing the generation of nonrelativistic net spin-polarized
currents (Fig.~5a)\cite{zhang2025crystal}. Previously reported altermagnets such as $\alpha$-MnTe and CrSb\cite{krempasky2024altermagnetic,lee2024broken,zeng2024observation,yang2025three} , were limited by \(C_3\) symmetry in each sublattice, which enforces isotropic conductivity and prevents the emergence of effective spin-polarized currents. MnTe$_2$ and RuO$_2$ either lacked spin conservation or suffered from low ordering temperatures and debated ground states\cite{zhu2024observation,feng2022anomalous,kiefer2025crystal,fedchenko2024observation}. In contrast, Rb$_{1-\delta}$V$_2$Te$_2$O simultaneously satisfies both symmetry and conductivity requirements for realizing nonrelativistic spin currents. Its room-temperature stability and layered nature further enable integration into vdW heterostructures, making it an ideal 2D spintronic platform\cite{lin2018structure,de2014ubiquitous,cao2018unconventional,flensberg2021engineered,zhao2020strain}.

\begin{figure}[htbp]
\centering
\includegraphics[width=\textwidth]{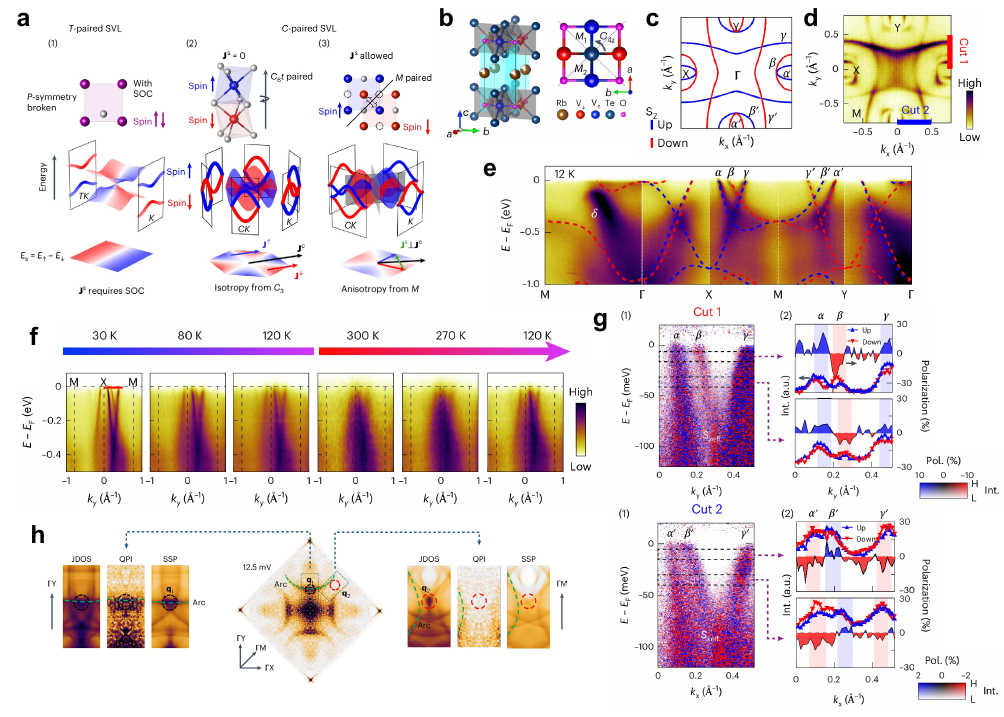} 
\caption{C-paired SVL and its experimental verification in layered altermagnet Rb$_{1-\delta}$V$_2$Te$_2$O\cite{zhang2025crystal}. a. Schematic illustration of three types of SVL. In T-paired SVL, spin splitting at opposite valleys arises from SOC. Type-I C-paired SVL features isotropic conductance for both spin channels, leading to non-polarized spin currents, while Type-II C-paired SVL exhibits anisotropic conductance and enables pure spin currents under an in-plane electric field. b. Crystal and magnetic structure of Rb$_{1-\delta}$V$_2$Te$_2$O. c,d. Calculated and ARPES-measured Fermi surfaces in the $k_x$–$k_y$ plane. e. ARPES intensity plot along $\Gamma$–$X$–$M$–$Y$–$\Gamma$ showing the main $\alpha$, $\beta$, and $\gamma$ bands, where the $\alpha$ and $\gamma$ branches carry spin polarization of one sign, and $\beta$ exhibits the opposite sign. 
The dashed red and blue curves represent calculated bands with opposite spin polarization. f. Temperature-dependent ARPES spectra measured along the $M$–$X$–$M$ direction from 30~K to 300~K. g. SARPES measurements taken along Cut~1 (top) and Cut~2 (bottom). Panels~(1) show the spin-resolved band dispersions, and panels~(2) display the extracted spin polarizations. h. QPI patterns from STS and corresponding JDOS and spin-dependent scattering probability (SSP) simulations.}
\end{figure}

The electronic structure (Figs.~5c–e) further corroborates SVL\cite{zhang2025crystal}. ARPES resolved three bands ($\alpha$, $\beta$, $\gamma$) near $E_F$, where $\alpha$ and $\gamma$ share spin orientation, opposite to $\beta$. The alternation of spin polarization between valleys matches first-principles results. Temperature-dependent measurements (Fig.~5f) confirmed the robustness of this spin splitting up to room temperature. While the bands were sharp and well-resolved at 30~K, the key spin-split features (e.g., $\alpha$ and $\gamma$ peaks) remained stable and discernible even at 300~K, demonstrating the high-temperature nature and reversibility of the SVL.
SARPES (Fig.~5g) directly verified the alternating spin polarization of C-paired SVL. In cut1, the $\alpha$ and $\gamma$ pockets showed $+S_z$ polarization, while the $\beta$ pocket showed $-S_z$; in the adjacent valley (cut2), these signs were fully reversed. In-plane ($S_x$, $S_y$) components were negligible, confirming an out-of-plane spin texture. These observations agree with theory and identify a $d$-wave spin texture in the altermagnetic state.

Finally, scanning tunnelling microscopy/spectroscopy (STM/STS) measurements (Fig.~5h) revealed quasi-particle interference (QPI) patterns consistent with spin-selective scattering\cite{zhang2025crystal,hoffman2002imaging,roushan2009topological}. Simulated joint density of states (JDOS) calculations yielded three main wavevectors: $\mathbf{q}_1$ (same-spin inter-valley scattering), $\mathbf{q}_2$ (opposite-spin inter-valley), and arc-like inter-$\gamma'$ scattering. Experimentally, $\mathbf{q}_2$ was absent, confirming suppression of opposite-spin scattering. The agreement between QPI, JDOS, and spin-dependent scattering probability simulations provides strong evidence for the distinctive C-paired SVL in Rb$_{1-\delta}$V$_2$Te$_2$O.

\subsection{MnTe: A domains-tunable altermagnet}

MnTe has emerged as a prototypical altermagnet in the hexagonal crystal system, characterized by a compensated antiparallel magnetic order accompanied by pronounced momentum-dependent spin splitting. In addition, its highly tunable magnetic domain textures and an experimentally established weak altermagnetic regime endow MnTe with remarkable versatility. This confluence of features has attracted intense interest, particularly for prospective applications in domain-wall-based spintronics and spin–orbitronics.

\begin{figure}[htbp]
\centering
\includegraphics[width=\textwidth]{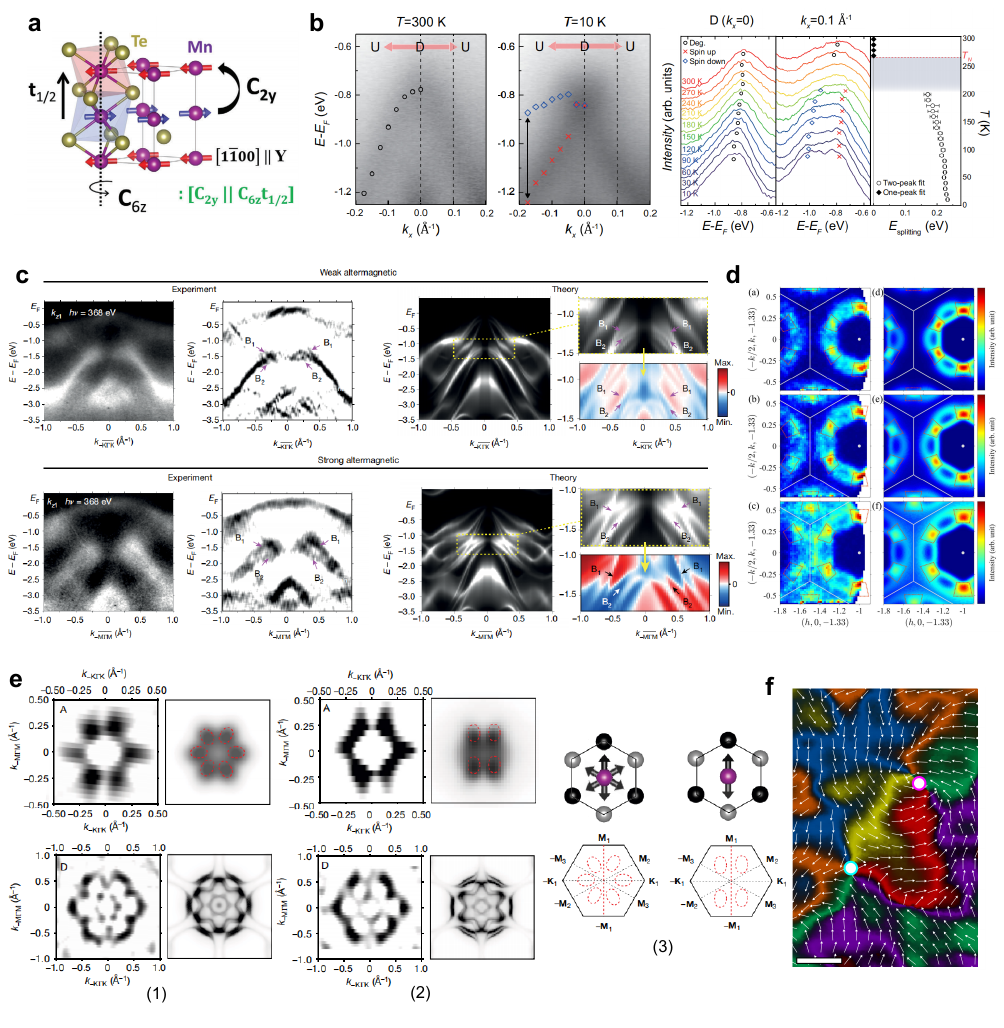}
\caption{
Altermagnetic characterization of MnTe\cite{liu2024chiral,lee2024broken,krempasky2024altermagnetic,amin2024nanoscale,hajlaoui2024temperature}. a. Crystal and magnetic structure of hexagonal $\alpha$-MnTe, in which opposing Mn spin sublattices are related by the non-symmorphic screw-axis operation $[C_{2y}\,||\,C_{6z}t_{1/2}]$. b. ARPES reveals the lifting of Kramers degeneracy across the magnetic transition ($T_N = 267$ K) in MnTe thin films, with temperature-dependent band splitting up to $\sim$370 meV. c. Soft-X-ray ARPES bands measured along $\Gamma$–$K$ and $\Gamma$–$M$ directions ($k_z = 0.35$ Å$^{-1}$), compared with calculations. Red and blue denote opposite spin polarizations, distinguishing weak and strong altermagnetic LKSD. d. Inelastic neutron scattering measurements reveal chiral magnon splitting with sixfold rotational symmetry and a $g$-wave intertwined energy contour in non-degenerate nodal planes. e. Experimental and theoretical constant-energy maps revealing the domain-dependent electronic symmetry in MnTe. (1) Multi-domain region exhibiting sixfold-symmetric contours. (2) Predominantly single-domain region where the Néel vector alignment leads to a pronounced twofold symmetry near the valence-band maximum, gradually recovering sixfold symmetry at higher binding energies. (3) Schematic comparison of multi-domain and single-domain configurations, showing sixfold and twofold electronic symmetries, respectively. f. Real-space visualization of altermagnetic domain textures by combined XMLD- and XMCD-PEEM imaging, showing vortex–antivortex pairs (magenta- and cyan-white circles) and the helical orientation of the order vector.
}
\end{figure}

Structurally, the stable phase of MnTe at room temperature is the hexagonal $\alpha$-MnTe, crystallizing in a nickel-arsenide–type structure with space group $P6_3/mmc$\cite{liu2024chiral,devaraj2024interplay}. Its magnetism arises entirely from Mn$^{2+}$ ions, and early studies identified $\alpha$-MnTe as a prototypical A-type antiferromagnet. Within the hexagonal symmetry, Mn sublattices couple ferromagnetically within each layer and antiferromagnetically between layers, forming a collinear magnetic structure that is antiparallel in real space but exhibits momentum-dependent alternating spin polarization (Fig.~6a).

Direct experimental evidence for such electronic reconstruction has been obtained by ARPES.
Lee \textit{et al.} observed the lifting of Kramers degeneracy across the magnetic transition ($T_N = 267~\mathrm{K}$) in MnTe thin films\cite{lee2024broken}, with a maximum spin splitting of about $370~\mathrm{meV}$. Temperature-dependent measurements\cite{lee2024broken, hajlaoui2024temperature, lee2025dichotomous} confirmed that the splitting emerges below $T_N$ and increases upon cooling (Fig.~6b).
Furthermore, J.~Krempaský \textit{et al.} employed soft X-ray ARPES to directly map the alternating spin texture\cite{krempasky2024altermagnetic}.
The lifting of Kramers spin degeneracy (LKSD) in MnTe represents a third mechanism beyond the conventional ferromagnetic and non-centrosymmetric origins, termed altermagnetic LKSD. It manifests in two regimes:
(i) a weak altermagnetic LKSD, in which the spin splitting is primarily induced by SOC within nodal planes (e.g., $k_z = 0$) while the system retains centrosymmetry; and
(ii) a strong altermagnetic LKSD, consistent with the core definition of altermagnetism, where the spin splitting is predominantly generated by nonrelativistic exchange interactions in regions away from the nodal planes, even in the complete absence of SOC. Their data (Fig.~6c) provides a decisive side-by-side comparison of the two LKSD regimes: the weak altermagnetic LKSD with a $\sim100~\mathrm{meV}$ splitting along the $\Gamma$–$K$ path within a nodal plane, and the strong altermagnetic LKSD with a splitting on the order of half an electron volt along the $\Gamma$–$M$ path away from the nodal plane\cite{krempasky2024altermagnetic}.
The direct observation of this massive, SOC-independent splitting along $\Gamma$–$M$ provides unambiguous experimental proof of strong, exchange-driven altermagnetism. From a spin-dynamics perspective, Liu \textit{et al.} used inelastic neutron scattering to reveal chiral magnon splitting featuring sixfold rotational symmetry and a $g$-wave-like intertwined energy contour in non-degenerate nodal planes (Fig.~6d), elucidating the altermagnetic coupling mechanism from a spin-wave standpoint\cite{liu2024chiral}.

The rich altermagnetic properties of MnTe are not limited to its bulk electronic structure; they extend to the mesoscopic scale, where the material exhibits highly tunable magnetic domain structures.
ARPES measurements by J.~Krempaský \textit{et al.} further revealed a direct correlation between the magnetic domain configuration and the electronic symmetry\cite{krempasky2024altermagnetic}.
In multi-domain regions, constant-energy maps exhibit sixfold symmetry reflecting the intrinsic hexagonal crystal symmetry, whereas in predominantly single-domain regions where the Néel vector aligns along one easy axis ($\Gamma$--$M_1$), the constant-energy contours near the valence-band maximum display a pronounced reduction to twofold symmetry (Fig.~6e). This symmetry lowering primarily originates from the selection of a single domain and becomes more evident at the valence-band edge due to the enhanced spin--orbit coupling associated with Te-$p$-derived states.
Early studies had already identified an easy axis along the $\langle 1\bar{1}00 \rangle$ direction and the existence of three controllable magnetic domain types.
The domain configuration can be set by field-cooling, where a strong magnetic field applied above $T_N$ and maintained during cooling forces the magnetic moments to align preferentially along field-correlated directions\cite{komatsubara1963magnetic, przezdziecka2008mnte, kriegner2017magnetic, hariki2024x-ray}.

Leveraging this tunability, J.~Amin \textit{et al.} achieved a breakthrough in the real-space visualization and manipulation of altermagnetism. By combining X-ray photoemission electron microscopy (PEEM) with XMLD—which is sensitive to the Néel order vector $\mathbf{L} = \mathbf{M}_1 - \mathbf{M}_2$—they unveiled micron-scale magnetic domains hosting nanoscale vortices and domain walls (Fig.~6f)\cite{amin2024nanoscale}. Through a combination of micro–nano fabrication and field-cooling protocols, controlled transitions from multi-domain to single-domain states were demonstrated, alongside the creation of distinct topological textures such as vortex–antivortex pairs and Bloch-type vortices in designed microstructures (e.g., hexagonal and triangular patterns). R.~Yamamoto \textit{et al.} recently employed transmission XMCD spectromicroscopy to directly visualize nanoscale altermagnetic domain structures in a 150–200~nm-thick lamella prepared from a bulk MnTe single crystal. Quantitative spectral analysis confirmed that this ordered state extends throughout the entire sample thickness, providing spectromicroscopic evidence for the intrinsic altermagnetic nature of bulk MnTe\cite{yamamoto2025altermagnetic}. Recent work also indicates that strain can be used to modulate the AHE in MnTe and to realize single magnetic domains in thin films, thereby further enriching its physical properties and underlying mechanisms\cite{liu2025strain,din2025unconventional}. This body of work not only provides the first real-space nanoscale view of magnetic textures in an altermagnet but also establishes a comprehensive methodology for imaging, controlling, and designing altermagnetic order.

\subsection{$g$-wave altermagnetism and Weyl topology in CrSb}
CrSb is a prototypical $g$-wave altermagnet that has attracted considerable attention as a model system for exploring the interplay between collinear altermagnetic order and topological electronic states. It features an exceptionally high Néel temperature of 705~K, a theoretically predicted nonrelativistic spin splitting of up to $\sim$1~eV near the Fermi level, and anticipated Weyl-type topological band crossings. These properties make CrSb an ideal platform for studying emergent phenomena at the intersection of magnetism and topology. A variety of experimental techniques, including X-ray scattering and ARPES~\cite{biniskos2025systematic,ding2024large,zeng2024observation}, have been employed to probe its fundamental electronic structure. In particular, ARPES directly visualizes the spin-split bands and reveals the microscopic origin and topological character of the altermagnetic state in CrSb.

CrSb crystallizes in a hexagonal NiAs-type structure (space group P6$_3$/mmc, Fig. 7a)\cite{zeng2024observation}. Its magnetic order arises from a collinear AFM configuration of Cr atoms, where adjacent spins align antiparallel, yielding zero net magnetization. However, unlike conventional AFM, the altermagnetic order in CrSb preserves the sixfold rotational symmetry while breaking the combined space–time ($\mathcal{PT}$) symmetry, resulting in opposite spin polarizations at +$k$ and –$k$ in momentum space. This momentum-dependent spin polarization forms a $g$-wave alternating pattern within the $ab$-plane, as shown in Fig. 7b\cite{ding2024large}, and theoretical calculations further predict a characteristic six-lobed spin-polarization distribution with periodic nodal planes along $k_z$ (e.g., $k_z$ = 0 and 0.5 c).

Since the latter half of 2024, multiple ARPES studies have directly observed non-relativistic spin splitting in CrSb, with magnitudes reaching up to $\sim$1~eV\cite{ding2024large,zeng2024observation,yang2025three}. One of the earliest reports, by Reimers \textit{et al.}\cite{reimers2024direct}, revealed clear spectroscopic signatures of band splitting in epitaxial CrSb thin films. Using both $P$- and $S$-polarized light, they verified the parity selection rules of the split bands, confirming their symmetry characteristics (Fig.~7c). Subsequently, Ding \textit{et al.}\cite{ding2024large,zeng2024observation} performed high-resolution ARPES on cleaved (0001) and (10$\bar{1}$0) surfaces of CrSb single crystals, observing preserved Kramers degeneracy along the high-symmetry $\Gamma$--$M$ direction but a pronounced $\sim$0.93~eV splitting along off-symmetry directions such as $P$--$D$ (Fig.~7d). The resulting Fermi-surface mapping displayed a sixfold $g$-wave modulation within the $ab$ plane and periodic $k_z$ dependence, consistent with theoretical predictions. This characteristic spin texture was further supported by circularly polarized resonant X-ray scattering with azimuthal scans (Biniskos \textit{et al.}\cite{biniskos2025systematic}), and complemented by SARPES measurements by Yang \textit{et al.}\cite{yang2025three}, which directly resolved the spin polarization and confirmed the $\sim$1~eV splitting near $E_F$ (Fig.~7e). To understand the microscopic origin of this giant $g$-wave splitting, Yang \textit{et al.}\cite{yang2025three} constructed a tight-binding model based on the experimental band dispersions. As illustrated in Fig.~7f, the splitting arises from third-nearest-neighbor hopping of Cr 3$d$ electrons mediated by Sb 5$p$ orbitals, forming an indirect Cr–Sb–Cr coupling pathway. This interaction breaks the translational spin-flip symmetry in the tight-binding framework, giving rise to a nonlocal spin polarization of about 1~eV near $E_F$. Moreover, this Sb-mediated coupling pathway is highly tunable by external strain or pressure, offering a route to engineer the magnitude of the altermagnetic splitting. Recent experimeents have demonstrated that by simultaneously applying strain to ultrathin (10nm) CrSb film\cite{li2025two}, the spin splitting can be completely eliminated, resulting in a unique spin-degenerate band structure\cite{https://doi.org/10.1002/adma.202515712}. 

The breaking of $\mathcal{PT}$ symmetry lifts the spin degeneracy and induces pronounced momentum-dependent band splitting, providing the essential symmetry condition for realizing nontrivial Weyl topology. Upon inclusion of SOC, the nodal rings predicted in its absence evolve into a finite number of discrete Weyl nodes, forming a topological band structure. ARPES measurements have directly observed these Weyl nodes along the $\Gamma$--$M$ direction and the associated surface Fermi arcs (Fig.~7g)\cite{li2025topological}. These observations establish CrSb as a prototypical material where altermagnetism and topological order coexist, offering a unique platform to explore the interplay between symmetry breaking, spin polarization, and topological electronic states\cite{li2025topological,lu2025signature}.

Beyond ARPES studies, transport measurements on CrSb have revealed equally rich physics. In single crystals, the Hall resistivity exhibits nonlinear field dependence dominated by ordinary Lorentz-force contributions, whereas epitaxial thin films show strain- and symmetry-dependent transport behaviors\cite{peng2025scaling,aota2025epitaxial}. Remarkably, by tuning the Néel vector orientation via epitaxial strain, researchers have realized a room-temperature anomalous Hall effect—highlighting the crucial role of symmetry control in governing the electronic and magnetic responses of altermagnets\cite{yu2025neel,zhou2025manipulation}.

\begin{figure}[htbp]
\centering
\includegraphics[width=\textwidth]{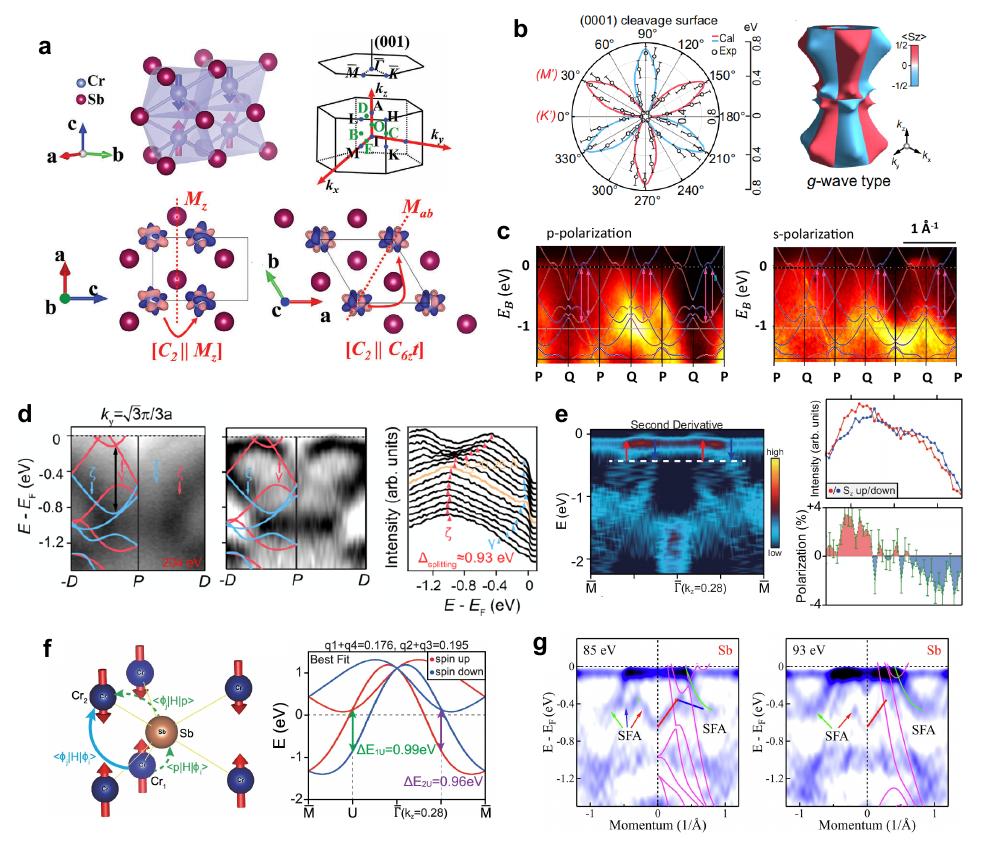} 
\caption{Altermagnetic band splitting and topological features of CrSb\cite{ding2024large,zeng2024observation,yang2025three,reimers2024direct,li2025topological}. a. Crystal and magnetic structure of CrSb.  
Cr moments form a collinear antiparallel alignment along the $c$-axis. The schematic also shows the corresponding Brillouin zone. b.Polar distribution of the maximum spin splitting magnitude ($\Delta_\mathrm{splitting}$) on the (0001) surface (left), showing a sixfold $g$-wave modulation with alternating spin polarizations.  
The 3D schematic (right) illustrates the corresponding $g$-wave-type spin texture in momentum space. c.The ARPES intensity and band structure calculations in high symmetry $P$–$Q$ path with $p$ and $s$-polarized photons. d.Observation of a giant nonrelativistic spin splitting ($\Delta_{\mathrm{split}} \approx 0.93$ eV) along the off-symmetry $P$–$D$ direction. e.SARPES along the $\Gamma$–$M$ direction at $h\nu$ = 80 eV. The second-derivative spectrum highlights band splitting near $E_F$, and spin-resolved MDCs reveal opposite spin polarizations for the split branches  f.Schematic and tight-binding analysis of the microscopic origin of the $g$-wave splitting in CrSb. g.ARPES evidence of Weyl nodes and surface Fermi arcs on the Sb-terminated (0001) surface of CrSb.\label{fig2}}
\end{figure}

\subsection{Noncoplanar Antiferromagnetism in MnTe$_2$}

By definition, AM requires a collinear magnetic order, where all local magnetic moments are strictly aligned either parallel or antiparallel to a common axis\cite{vsmejkal2022beyond, vsmejkal2022emerging,mazin2022altermagnetism,cheong2024altermagnetism}. In contrast, the ground-state magnetic structure of MnTe$_2$ (Fig.~8a) is a noncoplanar AFM: the local Mn moments possess components along multiple spatial directions, so that the spin projections ($S_x$, $S_y$, $S_z$) are simultaneously finite throughout momentum space and exhibit momentum-dependent variations\cite{zhu2024observation,wei2024crystal,chen2024emerging}. Importantly, despite this noncoplanar order, its band structure displays the key hallmark of altermagnetism: a large, non-relativistic spin splitting (persisting even without SOC) that alternates in sign across momentum space. 

From a magnetic-structure perspective, the noncollinear order in MnTe$_2$ is not an ad hoc assumption introduced to interpret ARPES data. It was established well before the concept of altermagnetism within a purely classical magnetic-exchange framework: isotropic exchange on the frustrated fcc lattice fixes the ordering wave vector but leaves the spin orientation degenerate, while additional classical interactions lift this degeneracy and select a robust noncollinear ground state, as confirmed by Mössbauer spectroscopy and symmetry-resolved single-crystal neutron diffraction\cite{hastings1970spin, 1977343, kasai1982mossbauer, burlet1997noncollinear}.

Fermi surface measurements (Fig.~8b) by Zhu \textit{et al.} provided momentum-space cuts that resolved the dispersive valence bands, showing good agreement with DFT calculations (Fig.~8c)\cite{zhu2024observation,zha2023improvement,schrunk2022emergence}. To distinguish this bulk spin texture from surface Rashba-type effects induced by SOC, photon-energy-dependent SARPES was performed (Figs.~8d and 8e)\cite{zhu2024observation}, revealing a systematic reversal of $S_x$ polarization along the $k_z$ direction. Specifically, polarization reversals between ``L'' and ``R'' points were observed at different photon energies, forming a plaid-like antisymmetric pattern in the $k_y=0$ mirror plane—providing clear evidence that the observed spin texture originates from the bulk antiferromagnetic order rather than surface-state effects. Temperature-dependent measurements (Fig.~8f) provided the final confirmation\cite{zhu2024observation}: a pronounced spin splitting was observed well below the Néel temperature ($T_N \approx 87$~K), which was strongly suppressed and vanished upon heating into the paramagnetic phase (e.g., above 110~K).

Thus, MnTe$_2$ provides a representative example of a noncoplanar antiferromagnet (AFM) that realizes momentum-dependent spin splitting analogous to theoretical predictions for collinear altermagnets. Its multicomponent spin texture forms a plaid-like alternating pattern throughout the Brillouin zone, expanding the scope of antiferromagnetic spintronics and illustrating that robust, non-relativistic spin polarization can be generated by unconventional magnetic symmetries beyond just the collinear altermagnetic case\cite{zhang2018spin,kimata2019magnetic,shindou2001orbital}.

\begin{figure}[htbp]
\centering
\includegraphics[width=\textwidth]{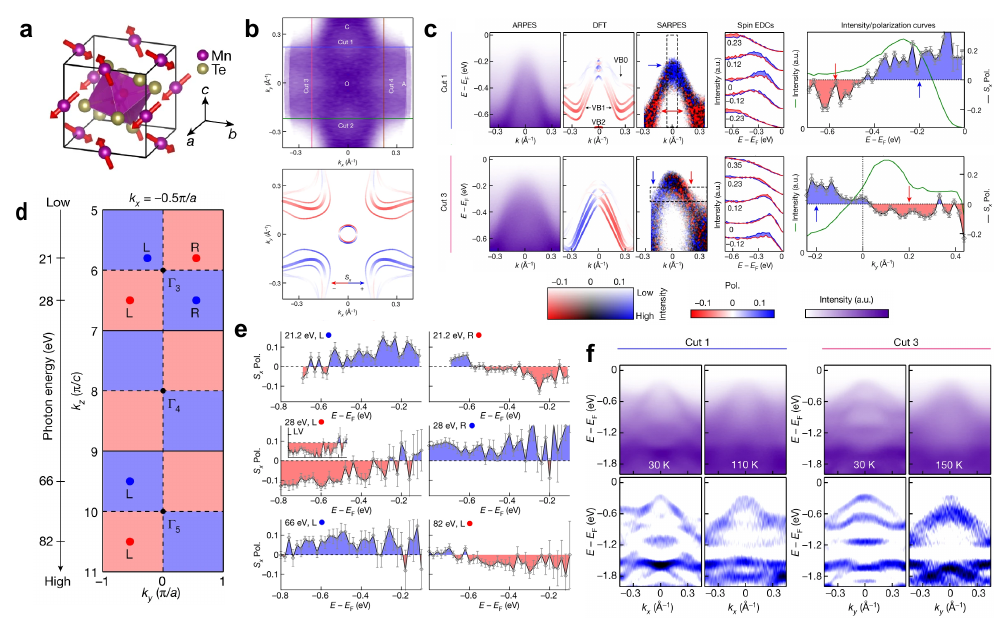} 
\caption{Electronic and spin texture of the noncoplanar antiferromagnet MnTe$_2$\cite{zhu2024observation}. a. Crystal structure of MnTe$_2$ showing the noncoplanar antiferromagnetic configuration, in which local Mn moments are oriented along multiple spatial directions. b. ARPES-measured Fermi surface in the $k_x$–$k_y$ plane, together with the corresponding calculated spin-resolved Fermi surface, revealing opposite spin polarizations. c. Comparison between ARPES, DFT, and SARPES results Cuts 1 and 3, showing the spin-split valence bands  and their $S_x$ polarization. d,e. Photon-energy-dependent SARPES measurements demonstrating systematic $S_x$-polarization reversals between the ``L'' and ``R'' points at $k_z = -0.2\pi/c$, $0.5\pi/c$, and $-0.5\pi/c$, forming a plaid-like antisymmetric pattern along the $k_z$ direction. f. Temperature-dependent ARPES band dispersions along Cuts 1 and 3, revealing pronounced spin splitting below $T_N$ = 87 K, which is strongly suppressed in the paramagnetic phase above 110 K.}
\end{figure}

\subsection{Emerging and prospective altermagnetic candidates and platforms}

Beyond the prototypical altermagnets discussed above, a growing number of materials and material platforms have been proposed as emerging or prospective realizations of altermagnetism, for which experimental evidence is still developing or remains partially incomplete. 
These systems nevertheless offer valuable opportunities to explore the diversity of altermagnetic phenomena, particularly in regimes where altermagnetism intertwines with electronic correlations, topology, or structural tunability. 
In the following, we first discuss several promising material candidates, and then outline forward-looking material platforms that are currently supported primarily by theoretical studies.

Among these candidates, layered CoNb$_4$Se$_8$ has recently attracted considerable attention as a particularly intriguing prospective altermagnet. It is theoretically predicted to host a $g$-wave altermagnetic state and, distinct from the previously discussed prototypical systems, features a layered van der Waals structure together with emergent correlated electronic orders. Although most existing studies are still at the preprint stage, a combination of ARPES, STM, and complementary probes has provided preliminary evidence for momentum-dependent nonrelativistic spin splitting as well as a coexisting density-wave instability~\cite{regmi2025altermagnetism,de2025optical,dale2024non,sakhya2025electronic,candelora2503discovery}. These early results position CoNb$_4$Se$_8$ as a promising platform for exploring the interplay between altermagnetism and electronic correlations, while further experimental and theoretical efforts remain essential to clarify its microscopic origin and magnetic ground state.

Structurally, CoNb$_4$Se$_8$ can be regarded as a derivative of the layered 2H-NbSe$_2$ system, in which Co atoms are intercalated in an ordered manner into one quarter of the octahedral sites within the van der Waals gaps. This ordered intercalation forms a well-defined 2$\times$2 superlattice and results in a centrosymmetric hexagonal structure with space group $P6_3/mmc$ (No.~194)~\cite{regmi2025altermagnetism}. Although CoNb$_4$Se$_8$ shares the same global centrosymmetric symmetry as its parent 2H-NbSe$_2$, the periodic Co sublattice breaks local inversion symmetry along the $c$-axis within the NbSe$_2$ layers, leading to alternating non-centrosymmetric environments on Co sites. When combined with the $A$-type antiferromagnetic order—ferromagnetic Co planes coupled antiferromagnetically along $c$—this configuration provides the structural and magnetic prerequisites for altermagnetism, enabling spin splitting without net magnetization. Magnetic susceptibility measurements reveal an antiferromagnetic transition at a Néel temperature ($T_N$) of approximately 168~K~\cite{regmi2025altermagnetism}, with the magnetic easy axis oriented along the crystallographic $c$-axis. Single-crystal neutron diffraction further confirms an \textit{A}-type antiferromagnetic structure, where Co magnetic moments align ferromagnetically within the \textit{ab}-planes and couple antiferromagnetically between adjacent planes along $c$.

Several independent studies have reported the hallmark signature of altermagnetism—a momentum-dependent, non-relativistic spin splitting (NRSS) of the electronic bands~\cite{de2025optical,dale2024non,sakhya2025electronic,candelora2503discovery}. 
ARPES measurements, which map different $k_z$ positions by varying the photon energy, reveal Fermi surfaces in excellent agreement with DFT calculations. 
At $k_z = 0.25$~\AA$^{-1}$, a distinct band doubling is observed, corresponding to the spin-up and spin-down branches of the split bands. 
A cut along the $\Gamma$–$M$ direction reveals a momentum separation of approximately 0.1~\AA$^{-1}$ between them. 
The sixfold rotational symmetry of this splitting and its periodic modulation along $k_z$ are consistent with a bulk $g$-wave symmetry~\cite{de2025optical}. 
Moreover, Dale \textit{et al.} employed spin- and angle-resolved electron reflection spectroscopy (spin-ARRES), revealing that this alternating spin texture persists in the unoccupied states above the Fermi level~\cite{dale2024non}. 
Together, these observations provide direct evidence for a bulk $g$-wave alternating magnetic symmetry.

Beyond its primary altermagnetic order, CoNb$_4$Se$_8$ exhibits a remarkably rich low-temperature phase diagram. 
STM measurements by Candelora \textit{et al.} revealed the emergence of a $2a_0 \times 2a_0$ charge–spin density wave below approximately 150~K~\cite{candelora2503discovery}. 
This observation is particularly significant, as such an electronic order is absent in the parent compound and not predicted by DFT calculations that consider only the ideal altermagnetic ground state, indicating a novel interplay between altermagnetism and electronic correlations. 
Consistent with this finding, ARPES measurements by both De~Vita \textit{et al.} and Candelora \textit{et al.} confirmed this transition, revealing folded replica bands that vanish above the same temperature ($\sim$150~K)~\cite{candelora2503discovery}. 
Furthermore, Candelora \textit{et al.} demonstrated that this emergent density-wave state can be tuned by an external magnetic field, while De~Vita \textit{et al.} showed that the entire low-temperature phase can be quenched by ultrafast optical excitation—realizing the first ``optical switching’’ of an altermagnetic state~\cite{de2025optical,candelora2503discovery}.

The discovery of altermagnetism in CoNb$_4$Se$_8$ represents more than the identification of a new material—it establishes a uniquely versatile experimental platform. 
Owing to its layered van der Waals structure, CoNb$_4$Se$_8$ offers a promising pathway toward two-dimensional spintronics and heterostructure-based device architectures, providing a key advantage over many bulk altermagnets. 
Notably, it is the first reported material in which a tunable correlated electronic order—a density wave—appears to coexist with altermagnetism, offering new insights into the interplay between spin-split band structures and many-body interactions. 
As studies continue to confirm its $g$-wave altermagnetic character, uncovering the microscopic coupling between this magnetic order and the emergent density-wave phase remains a crucial and intriguing direction for future investigation.

In addition to CoNb$_4$Se$_8$, several other materials have been proposed as emerging altermagnetic candidates, although comprehensive ARPES verification remains limited or indirect.
Jadupati~Nag~\textit{et al.} reported the first noncentrosymmetric ($I4_1md$, No.~109) collinear antiferromagnetic Weyl semimetal GdAlSi~\cite{PhysRevB.110.224436}, exhibiting NRSS in its AFM ground state. Combining DFT calculations, VUV-ARPES, and magneto-transport measurements, they revealed the interplay between altermagnetism and topology in this compound. As shown in Fig.~9a~\cite{PhysRevB.110.224436}, the (001)-surface VUV-ARPES map displays two concentric electron pockets around $\Gamma$ and arc-like surface states along $\Gamma$--$X$, consistent with the theoretically predicted spin-split surface states induced by bulk altermagnetism. Given the high surface sensitivity of VUV-ARPES, the observed splitting originates mainly from surface bands, while direct evidence of bulk spin splitting remains unresolved. Future soft-X-ray ARPES and resonant magnetic X-ray diffraction experiments will be required to confirm the coexistence of bulk altermagnetism and Weyl topology in GdAlSi.

Another candidate is Mn$_5$Si$_3$, a hexagonal compound with the $P6_3$/$mcm$ structure\cite{reichlova2024observation}. Its distinctive electronic state arises from the interplay between in-plane $C_2$ rotational symmetry and time-reversal operation, which enforces alternating spin orientations perpendicular to the propagation vector and produces a $d$-wave spin texture in momentum space(Fig.~9b). The pronounced spin splitting in Mn$_5$Si$_3$ gives rise to robust, electrically detectable transport responses. Epitaxial films exhibit a spontaneous AHE in the absence of external fields\cite{reichlova2024observation}, evidencing altermagnetism-induced spin polarization, along with a sizable anomalous Nernst effect\cite{badura2025observation} (ANE, $\sim$0.26 $\mu$V/K) reflecting momentum-dependent spin textures. Furthermore, SOT enables reversible 180° electrical switching of the Néel vector under a small assisting field ($\sim$0.2 kOe), with the magnetic state stably read out via AHE—underscoring Mn$_5$Si$_3$ as a promising platform for low-power antiferromagnetic spintronics\cite{reichlova2024observation}.

Ca$_3$Ru$_2$O$_7$ represents another promising candidate for altermagnetism. Theoretically, its $A$-type antiferromagnetic ground state is predicted to evolve into a $P$-$2$ $d$-wave altermagnetic phase under strain\cite{leon2025hybrid}, where orbital-selective spin splitting emerges from the interplay between nonrelativistic exchange and relativistic SOC. Recent ARPES measurements revealed anomalous band dispersions and clear Fermi-surface reconstruction in Ca$_3$Ru$_2$O$_7$\cite{horio2021electronic,markovic2020electronically}, but no discernible spin splitting or other fingerprints of altermagnetism have been observed. The definitive experimental confirmation of such an order therefore remains outstanding to date. The unique crystal symmetry of Ca$_3$Ru$_2$O$_7$ (space group $Bb2_1m$) is predicted to enable strain-tunable spin splitting. When compressive strain exceeds 0.02, enhanced hybridization between Ru-$d_{xy}$ and O–$p$ orbitals is thought to stabilize the altermagnetic phase. Moreover, the lack of inversion symmetry gives rise to a finite Berry curvature and spin-momentum locking, positioning Ca$_3$Ru$_2$O$_7$ as an ideal platform for exploring nonequilibrium quantum phenomena such as elasto-Hall effects and strain-controlled spin transport. Notably, experiments have demonstrated that compressive strain drives a bandwidth-mediated spin-reorientation transition (SRT) from AFM$_a$ (antiferromagnetic phase A) to AFM$_b$ via octahedral distortion-modulated hopping (Fig.~9c)\cite{dashwood2023strain}. This dual sensitivity to strain—a predicted AM phase and an observed SRT—positions Ca$_3$Ru$_2$O$_7$ as a frontier system bridging theoretical AM predictions with low-power spintronic applications.

\begin{figure}[htbp]
\centering
\includegraphics[width=\textwidth]{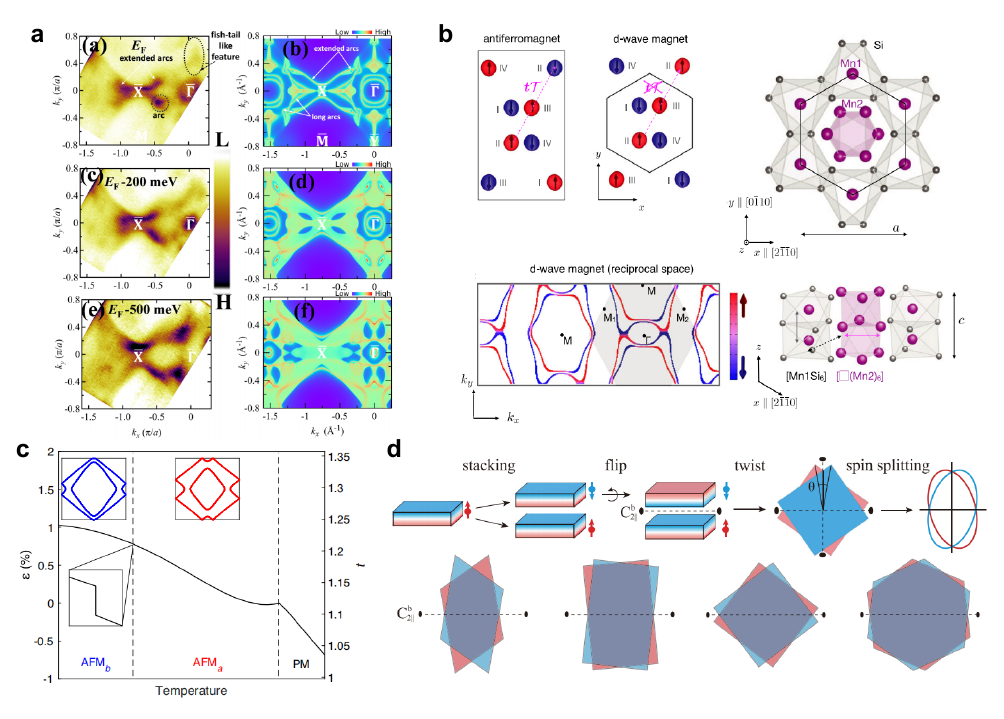} 
\caption{Other unverified altermagnetic candidates\cite{PhysRevB.110.224436,liu2024twisted,reichlova2024observation,dashwood2023strain}. a. Electronic structure of the GdAlSi (001) surface. (a), (c), (e) Experimentally measured Fermi surface and constant energy contours at the Fermi level $E_F$, $E_F$$\mathrm{-200}$ meV, and $E_F$$\mathrm{-500}$ meV, respectively. (b), (d), (f) Corresponding first-principles calculation results. b. Unconventional \textit{d}-wave magnetism in Mn$_5$Si$_3$. c. Strain-coupled spin reorientation in Ca$_3$Ru$_2$O$_7$, where the temperature-dependent internal strain $\varepsilon$ reveals a bandwidth-driven SRT accompanied by slight lattice distortion and distinct Fermi-surface features between the two magnetic phases.  d. Two-dimensional altermagnetism in twisted van der Waals bilayers.}
\end{figure}

Beyond specific material candidates, altermagnetism has also been theoretically predicted in broader material platforms, most notably twisted magnetic van der Waals (MvdW) bilayers. Recent theoretical studies have revealed that MvdW bilayers derived from two-dimensional Bravais lattices universally possess an in-plane $C_2$ rotational symmetry, which enables altermagnetic spin splitting without net magnetization\cite{liu2024twisted} (Fig.~9d). In such systems, the emergence of altermagnetism is dictated primarily by symmetry principles rather than specific microscopic details, providing a conceptually general and highly tunable route toward two-dimensional altermagnetism.

Building on this symmetry framework, twisting the bilayer offers a robust means to induce and continuously tune altermagnetic order across a wide range of MvdW magnets. This mechanism is independent of the monolayer symmetry—applicable to systems such as CrI$_3$, MnBi$_2$Te$_4$, and FeTe—and remains robust against interlayer sliding\cite{feng2021antiferromagnetic,shen2022high}. Moreover, it allows the realization of distinct $d$-, $g$-, and $i$-wave altermagnetic states through control parameters including twist angle, doping, strain, and pressure. For example, in the transition-metal oxyhalide VOBr, theoretical calculations predict that tuning the twist angle and Fermi level can generate a giant spin Hall angle far exceeding existing experimental reports\cite{miao20182d}. The resulting anisotropic, nonrelativistic spin splitting is in principle directly accessible by ARPES and SARPES measurements\cite{lee2024broken,zhu2024observation,krempasky2024altermagnetic}, while twist-dependent spin currents can be probed via spin-torque ferromagnetic resonance (ST-FMR)\cite{liu2011spin,husain2020large}. At present, these twisted-layer realizations are supported primarily by theoretical studies and are therefore best regarded as forward-looking material platforms.

This review centers on ARPES-based investigations of altermagnets, emphasizing representative systems that have been experimentally established thus far. Nevertheless, a broad range of promising candidates and material platforms—spanning both emerging compounds and symmetry-driven architectures—remain to be explored and experimentally verified. A comprehensive compilation of such potential altermagnetic systems is available in Ref.~\cite{gao2025ai}.

\section{Outlook}\label{sec4}

The rapid progress of altermagnetism has reshaped our understanding of spin–momentum coupling and stimulated new perspectives in quantum magnetism. Despite these advances, fully characterizing the momentum-dependent spin splitting and definitively correlating it with real-space magnetic domain structures remain major challenges, demanding advanced spectroscopic and imaging techniques that combine high spatial and momentum resolution. Integrating micro-beam ARPES with \textit{in-situ} STM, for example, will be essential for simultaneously visualizing band dispersions in momentum space and spin textures in real space, directly bridging these two crucial perspectives of altermagnetic order. Meanwhile, strain engineering, magnetic-domain control, and heterostructure design provide powerful routes to tune nonrelativistic spin splitting and to stabilize new functional quantum phases. Beyond altermagnetism, the exploration of unconventional magnetic states—ranging from odd-parity ($p$-wave) magnets to complex higher-order spin-ordered systems—marks an emerging frontier, deepening our understanding of how symmetry and electronic correlations give rise to novel magnetic and topological phenomena. Looking forward, \textit{in-situ} high-resolution ARPES, nanoscale spin imaging, and symmetry-guided material design will be the essential pillars for revealing the microscopic origins and technological potential of this new generation of quantum magnetic materials.

\subsection{ARPES combined with micro-beam or \textit{in-situ} STM}

ARPES provides direct experimental access to nonrelativistic spin splitting in altermagnets. However, the spatial averaging over magnetic domains often causes signal cancellation or blurring, which obscures the intrinsic momentum-dependent spin polarization. Overcoming this limitation highlights the urgent need for advanced high-spatial-resolution techniques, such as micro-beam ARPES, capable of resolving domain-specific electronic structures.

On the one hand, micro-beam ARPES focuses the probe to the micrometer or even nanometer scale \cite{miyai2024visualization, schneider2012expanding}, enabling site-specific measurements on individual magnetic domains and effectively suppressing signal blurring caused by multi-domain averaging. Recently developed spectrometers that integrate advanced designs such as a five-electrode electron mirror corrector have achieved spatial resolutions approaching 1~nm \cite{yang2025a}, opening the way to directly resolve momentum-dependent spin splitting in altermagnets at the nanoscale.

On the other hand, STM and its spin-resolved variant (sp-STM) provide a powerful route for real-space imaging of magnetic domain structures with atomic precision \cite{wiesendanger2009spin}. sp-STM combines the sub-angstrom spatial resolution of STM with the magnetic sensitivity of spin-polarized tunneling, enabling direct visualization of surface spin textures and magnetic anisotropy at the atomic scale. Recent studies have demonstrated that sp-STM can achieve atomic-resolution magnetic imaging of skyrmions in Fe/Ir(111) via spin-polarized resonance states (Fig.~10a)\cite{schlenhoff2020real-space}. It has also directly resolved helical magnetic domains and domain-wall atomic structures in MnGe, revealing complex topological spin textures such as “target-like” and “$\pi$-like” patterns, along with reversible manipulation of their cores (Fig.~10b)\cite{repicky2021jacob}.

Combining ARPES with STM establishes a powerful experimental framework that directly links momentum-resolved electronic structures with real-space manifestations of collective electronic orders\cite{yang2022large-gap, meng2023thickness-dependent}, as exemplified in Rb$_{1-\delta}$V$_2$Te$_2$O, where STM plays a key role in identifying the electronic signatures of altermagnetism. In 2H-TaS$_2$, this combined approach reveals a strong-coupling CDW, with STM resolving a large local gap and ARPES showing that the CDW gap is mainly distributed on the $K$-centered Fermi-surface barrels, accompanied by pronounced particle--hole asymmetry and pseudogap-like behavior (Fig.~10c)\cite{zhao2017orbital, wang1990energy}. Related studies on 2H-Na$_x$TaS$_2$ further point to a Fermi-patch–driven CDW mechanism beyond conventional Fermi-surface nesting\cite{shen2007novel}. These results highlight the essential role of multi-modal spectroscopic approaches in disentangling the microscopic mechanisms underlying emergent ordered phases in low-dimensional quantum materials.

Other microscale characterization techniques—including micro-focused X-ray magnetic circular dichroism, electron holography, and NV-center-based magnetic force microscopy—can also resolve magnetic domain structures from sub-micron to nanometer scales \cite{zhao2025an}. These high-spatial-resolution approaches are highly complementary, together forming an essential experimental pathway for visualizing the coupled momentum- and real-space spin textures that define altermagnetic materials.

\begin{figure}[H]
\centering
\includegraphics[width=\textwidth]{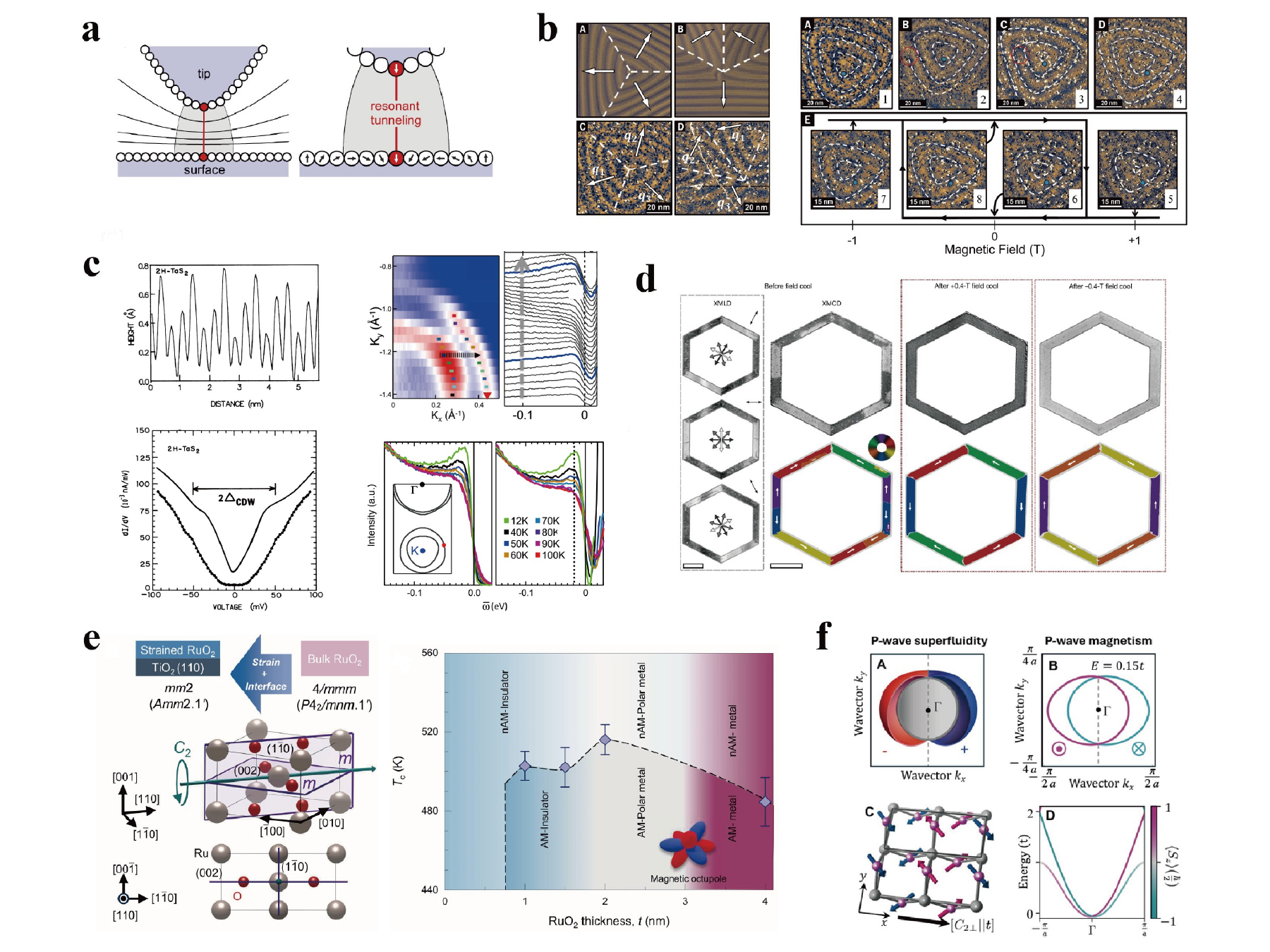} 
\caption{a. Schematic of resonant tunneling (left: field emission, right: resonant tunneling) \cite{schlenhoff2020real-space}. This mechanism enables atomic-scale spin-resolved imaging via tunneling current localized to the atomic-scale channel. b. SP-STM imaging and manipulation of target and $\pi$-state spin textures in MnGe thin films \cite{repicky2021jacob}. Left: Micromagnetic simulations (A, B) and experimental observations (C, D) of the target and $\pi$-state structures; Right: Reversible topological evolution and hysteretic behavior of the target state under current pulsing (A--D) and applied magnetic field (E). 
c. Synergistic ARPES and STM investigation of bulk 2H-TaS$_2$\cite{zhao2017orbital,wang1990energy}.
Top left: STM line profile resolving the atomic lattice together with the CDW modulation.
Bottom left: Quantitative determination of the CDW gap.
Top right: Fermi-surface map and corresponding EDCs along the indicated momentum cut.
Bottom right: Temperature-dependent EDCs measured at the red momentum point.
d. Field-controlled altermagnetic single domains in MnTe \cite{amin2024nanoscale}. Field cooling at $\pm$0.4 T converts multi-domain states (Left) into single domains (Right), with the L-vector switched by the field sign. e. Strain engineering of altermagnetism in RuO$_2$ thin films \cite{jeong2024altermagnetic}. Strain tunes the altermagnetic stiffness by lattice distortion (Left) and induces a novel polar metallic phase (Right), demonstrating the ability to control and induce new states. f. Unconventional $p$-wave phases\cite{hellenes2023p}. (A) $p$-wave superfluidity in $^3$He, showing a parity-breaking Cooper-pairing gap with opposite spin orientations across the Fermi surface. (B) Schematic of the corresponding $p$-wave magnetic state, featuring parity-breaking, spin-polarized Fermi surfaces with opposite spin helicities. (C) Real-space representation of a coplanar noncollinear spin texture that realizes this $p$-wave magnetic order. (D) Calculated spin-polarized band structure based on the model in (B, C).}
\end{figure}

\subsection{Strain or magnetic domain engineering in altermagnetic materials}

Magnetic domain engineering provides an effective route for controlling altermagnetic order. By constructing single-domain regions, it can suppress signal averaging caused by multi-domain sampling and enable nanoscale manipulation of the altermagnetic order vector \textbf{L} in real space. In MnTe, Amin O. J. \textit{et al.} achieved precise domain control through micro/nanofabrication combined with field-cooling (Fig.~10d)\cite{amin2024nanoscale}. Their results revealed that micrometer-scale geometry strongly dictates the domain configuration, while field cooling further stabilizes single-domain regions, offering a practical pathway toward high-fidelity, programmable altermagnetic logic units.

Beyond domain engineering, strain engineering has emerged as another highly effective and versatile approach for controlling the physical properties of altermagnets. By modifying lattice symmetry, strain can directly tune the defining characteristic of altermagnetism—the non-relativistic spin splitting. In RuO$_2$, Jeong et al. revealed that this modulation originates from adjustments in the anisotropic altermagnetic stiffness(AAS) within a phenomenological framework\cite{jeong2024altermagnetic}. Their experiments on ultrathin epitaxial RuO$_2$ films confirmed this mechanism, showing that strain not only induces high-temperature altermagnetism but also stabilizes a polar metallic phase (Fig.~10e)\cite{jeong2024altermagnetic}. Moreover, systematic investigations across materials such as MnF$_2$, KV$_2$Se$_2$O, MnTe, and CrSb have demonstrated that external stress can effectively tailor magnetic properties—ranging from the magnetic transition temperature to the magnitude of altermagnetic spin splitting \cite{li2025magnetic, fan2025high-pressure, devaraj2024interplay, carlisle2025tuning, zhou2024Crystal-design, zhou2025manipulation, reichlova2024observation, han2024electrical, sim2025leveraging, leon2025hybrid, chen2025strain-modulated}.

Looking ahead, the synergistic integration of magnetic domain and strain engineering will open new research directions. For instance, introducing localized strain fields in heterostructures to define magnetic domain pathways, or utilizing electric-field-controlled strain (e.g., via piezoelectric substrates) for low-power magnetic state switching, holds promise for developing high-speed, high-density, non-volatile information storage and processing devices based on altermagnetism.

\subsection{Unconventional magnets}
Beyond altermagnetism lies a broader classification of matter—unconventional magnets—representing an emerging class of quantum states. In this hierarchy, altermagnets correspond to even-partial-wave unconventional magnets (e.g., \textit{d}- and \textit{g}-wave)\cite{vsmejkal2022emerging}, whereas their odd-parity counterparts are exemplified by \textit{p}-wave magnets (and theoretically, higher-order \textit{h}-wave types with complex momentum responses)\cite{ezawa2025third,yu2025odd}. Odd-parity magnets are characterized by sign-reversing spin polarization under spatial inversion (Fig.~10f), conceptually analogous to \textit{p}-wave superconducting pairing and orbital symmetries~\cite{hellenes2023p,yu2025odd}. Experimentally, NiI$_2$ has been identified as the first realization of \textit{p}-wave magnetism~\cite{song2025electrical}, representing an insulating \textit{p}-wave magnetic state. More recently, the metallic compound Gd$_3$(Ru$_{1-x}$Rh$_x$)$_4$Al$_{12}$ ($x \approx 0.05$) has exhibited \textit{p}-wave-like anisotropic electronic behavior and a giant anomalous Hall effect~\cite{yamada2025gapping,yamada2025metallic}, suggesting unintended $\mathcal{T}t$-symmetry breaking that challenges the ideal parity-only \textit{p}-wave condition. Its pronounced momentum-dependent spin splitting further implies that interfacing \textit{p}-wave magnets with conventional \textit{s}-wave superconductors could induce spin-triplet pairing with finite center-of-mass momentum, potentially realizing topological superconductivity accompanied by 0–$\pi$ Andreev bound states and Majorana zero modes~\cite{cai2015coexistence,sukhachov2025coexistence,hellenes2023p}. Such \textit{p}-wave magnet/\textit{s}-wave superconductor heterostructures thus constitute ideal quantum platforms for exploring electrically tunable spin textures and emergent topological superconductivity. Recent preprints on CeNiAsO further report anisotropic resistivity consistent with \textit{p}-wave predictions~\cite{zhou2025anisotropic}, underscoring that odd-parity magnetism is rapidly emerging as a frontier in correlated, spin–orbit–free quantum materials.

While the \textit{d}/\textit{g}/\textit{p}-wave classification scheme is largely rooted in collinear or coplanar magnetic orders, theoretical frameworks proposed by C.~Wu and S.~C.~Zhang~\cite{PhysRevLett.93.036403,PhysRevB.75.115103,PhysRevB.80.104438} indicate that a vast range of noncoplanar magnetic configurations remain unexplored. The concept of unconventional magnetism thus constitutes a broad and diverse research field. This progression—from early Fermi-liquid–based models to the recent group-theoretical discovery of altermagnets—deepens our understanding of how symmetry and electronic correlations are coupled in quantum magnetic systems. Looking ahead, additional classes of magnetic order are likely to emerge, further refining the taxonomy of unconventional magnets and bridging distinct paradigms of modern magnetism within a unified theoretical framework.

\section{Conclusion}\label{sec5}

The rapid development of ARPES has transformed altermagnetism from a purely symmetry-based concept into a directly verifiable band phenomenon. By combining conventional ARPES, SARPES, CD-ARPES, soft X-ray ARPES, and micro/nano-beam spatially resolved measurements, it has become possible to directly visualize the characteristic nonrelativistic, momentum-dependent spin splitting and its $d$-, $g$-, or $i$-wave angular symmetries in altermagnets. In representative systems, ARPES studies on KV$_2$Se$_2$O, Rb$_{1-\delta}$V$_2$Te$_2$O, MnTe, CrSb and MnTe$_2$ have resolved distinct spin textures and nonrelativistic band splittings, while RuO$_2$ remains a debated case where clear band splitting has not been observed and the intrinsic magnetic ground state continues to be under scrutiny. Collectively, these systems demonstrate that ARPES not only visualizes spin splitting but also distinguishes bulk and surface states, disentangles spin–orbit– and exchange-driven mechanisms, and elucidates the coupling between spin texture, crystal symmetry, and magnetic-domain structure—thereby clarifying experimental controversies and uncovering connections to topological and correlated electronic phenomena. Looking forward, ARPES techniques with domain-resolved capabilities (single-domain selection, \textit{in-situ} strain or twist control), in combination with STM/QPI and resonant X-ray scattering, will provide decisive evidence for resolving outstanding controversies and verifying new candidate materials. Consequently, ARPES has evolved beyond a mere spectroscopic probe into an indispensable tool for altermagnetism research—establishing the foundation for verifying band, spin, and symmetry properties in future candidates and for designing spintronic functionalities based on altermagnetic order.

\bmhead{Acknowledgements}

This work is supported by National Key R\&D Program of China (Grants No. 2024YFA1408103, No. 2023YFA1406304) and National Natural Science Foundation of China (Grant No.12494593). D.W.S. acknowledges the Anhui Provincial Natural Science Foundation (No. 2408085J003). We thank the Shanghai Synchrotron Radiation Facility (SSRF) for the beamtime on beamline 03U (31124.02.SSRF.BL03U), which is supported by ME$^2$ project under contract No. 11227902 from National Natural Science Foundation of China.

\bmhead{Author contributions}

J.Y.L prepared the manuscript content and contributed to its writing and revision. X.M and X.N.Z drafted and reviewed specific sections. W.C.J assisted in manuscript writing. D.W.S supervised the research, and provided critical revision and final checking of the manuscript.

\bmhead{Availability of data and materials}

Not applicable.

\bibliography{references}

\end{document}